\documentclass[a4paper,fleqn]{cas-sc}

\usepackage[authoryear]{natbib}

\newcommand{\Bb}{{\boldsymbol{\mathnormal b}}}

\newcommand{\Be}{{\boldsymbol{\mathnormal e}}}

\newcommand{\Bg}{{\boldsymbol{\mathnormal g}}}

\newcommand{\Bk}{{\boldsymbol{\mathnormal k}}}

\newcommand{\Bn}{{\boldsymbol{\mathnormal n}}}

\newcommand{\Br}{{\boldsymbol{\mathnormal r}}}
\newcommand{\Bs}{{\pmb{\mathnormal s}}}
\newcommand{\Bt}{{\boldsymbol{\mathnormal t}}}

 \newcommand{\Bsigma} {\ensuremath{\boldsymbol\sigma}}

 \newcommand{\dx}{\text{\rm d}x}
 \newcommand{\dy}{\text{\rm d}y}
 
 \newcommand{\ds}{\text{\rm d}s}
 \newcommand{\dr}{\text{\rm d}r}

\begin{document}
\let\WriteBookmarks\relax
\def\floatpagepagefraction{1}
\def\textpagefraction{.001}
\shorttitle{Dislocations in random stress fields}
\shortauthors{M. Zaiser and R. Wu}

\title [mode = title]{Dislocations in the elastic fields of randomly distributed defects}                      

\author[1]{Ronghai Wu}[style=chinese]
\affiliation[1]{organization={State Key Lab of Solidification Processing, School of Materials Science and Engineering, Northwestern Polytechnical University},
                postcode={{710072}}, 
                postcodesep={}, 
                city={Xi'an},
                country={P.R. China}}

\author[2]{Michael Zaiser}[type=editor,
                        auid=000,bioid=1,
                        orcid=0000-0001-7695-0350,]
\cormark[1]
\ead{michael.zaiser@fau.de}

\affiliation[2]{organization={Department of Materials Science, WW8-Materials Simulation, Friedrich-Alexander Universität Erlangen-Nürnberg},
                addressline={Dr.-Mack-Str. 77}, 
                city={F\"urth},
                postcode={90762}, 
                country={Germany}}
\cortext[cor1]{Corresponding author}

\begin{abstract}
In recent years, the behavior of dislocations in random solid solutions has received renewed interest, and several models have been discussed where random alloys are treated as effective media containing random distributions of dilatation and compression centers. More generally speaking, the arrangement of defects in metals and alloys is always characterized by statistical disorder, and the same is true for the fluctuating fields that arise from the superposition of the stress and strain fields of many defects. In order to develop models for the dynamics of dislocations interacting with such fields, a statistical description of the dislocation energy landscape and the associated configurational forces is needed, as well as methods for coarse graining these forces over dislocation segments of varying length and shape. In this context the problem arises how to regularize the highly singular stress fields associated with dislocations and other defects. Here we formulate an approach which is based upon evaluating the interaction energies and interaction forces between singular dislocations and other defects including solutes, modelled as point-like dilatation centers, and other dislocations. We characterize the interactions in terms of the probability densities of interaction energies and interaction forces, and the corresponding spatial correlation functions. We also consider the effects of dislocation core regularization,  either in terms of continuously distributed Burgers vectors or by using the formalism of gradient elasticity of Helmholtz type to formulate a regularized energy functional. 
\end{abstract}

\begin{highlights}
\item A unified approach for deriving the probability density functions and spatial correlation functions of internal fields created by randomly distributed defects.
\item Proof that the classical assumption of short-range correlated Gaussian random fields is not adequate for stresses, displacements or energy densities created by randomly distributed defects.
\item Equations for the interaction of dislocations with randomly distributed solutes or forest dislocations are derived
\item Equations for the resolved shear stress field of disordered dislocation systems are validated by 3D discrete dislocation dynamics.
\end{highlights}

\begin{keywords}
dislocations \sep defect interactions \sep random stress fields
\end{keywords}
\maketitle

\section[Introduction]{Introduction}

The motion of dislocations is controlled by their interactions with other defects, who are generally arranged in a statistically disordered manner. Such interactions include atomic defects as in solid solutions (including so-called high entropy alloys where 'every atom is a solute'), which govern solute hardening, but also extended defects such as other dislocations, which give rise to forest hardening. The currently renewed interest in solute hardening has led to a number of recent works which aim at a statistical description of the elastic fields created by randomly distributed defects, in view of enabling dislocation dynamics simulations withoiut the necessity to explicitly resolve every dislocation-defect interaction. Several authors  (e.g.\citet{Varvenne2016theory,Varvenne2017solute,larosa2019solid}) have argued that dislocation-solute interactions are mainly governed by the stress fields caused by the atomic misfit fluctuations that emerge as atoms of different size are arranged on a common lattice, forming an uncorrelated random arrangement of dilatation/compression centers. Recently, \citet{geslin2021microelasticityI,geslin2021microelasticityII} evaluated the magnitude and spatial correlations of the ensuing random stress/strain fields
and used this information to analyze the motion of dislocations in a correlated random stress field mimicking the influence of randomly distributed point defects on dislocation motion \citep{rida2022influence}. 

Similar approaches have also been proposed for dealing with dislocation-dislocation interactions. In the context of idealized, two-dimensional dislocation systems (i.e., systems of straight parallel dislocations), \citet{groma1998probability} evaluated the stress probability density function arising from superposition of the stress fields of randomly arranged dislocations, and used this information to drive a stochastic, two-dimensional dislocation dynamics simulation \citep{groma2000dislocation}. Results have also been presented for two-dimensional walls in which dislocations are randomly arranged - a situation which might, for example, apply to the stress field of geometrically necessary dislocations immobilized at a planar boundary. The random elastic field of such 2D planar random dislocation arrays was characterized by \citet{saada1995long} and \citet{zaiser2002dislocation}, who demonstrated that the results differ radically from those obtained for walls with regular equi-spaced dislocations. 

In the present investigation we adopt a unifying approach to evaluating the probability density functions and spatial correlation functions of three-dimensional systems of defects, for the case of point defects that can be modelled as compression/dilatation centers and for systems of dislocations. 
Borrowing from methods developed in the theory of electrolytes and plasmas \citep{holtsmark1919verbreiterung}, and in cosmology \citep{chandrasekhar1943stochastic}, we consider the fields arising from superposition of elastic fields of singular defects as well as of their core-regularized counterparts. We first focus on the evaluation of correlation functions in Section \ref{sec:2}, and then move to calculations of probability density functions in Section \ref{sec:3}

\section{Random fields created by lattice defects: Correlation funtions}
\label{sec:2} 

\subsection{Point defects: Interaction of a dislocation with a random arrangement of compression or dilatation centers} 

\subsubsection{Interaction of a dislocation with a single dilatation center, singular dislocation}

\paragraph{Interaction energy}

We first consider the interaction between a dislocation and a single defect which whithout loss of generality we take to be a dilatation center. We start by evaluating the interaction energy, building our treatment on the classical result that the energy of a dilatation center of strength $v$ located at $\Br$ in a generic stress field $\Bsigma(\Br)$ is simply the hydrostatic pressure at the location of the dilatation center, multiplied with the strength of the center \cite{eshelby1956continuum}, 
\begin{equation}
	E(\Br) = \frac{v {\rm Tr}\Bsigma(\Br)}{3}. 
\end{equation}
This result holds equally for classical elasticity and for gradient elasticity theories of the Helmholtz or Bi-Helmholtz types \citep{lazar2019non}. We consider without loss of generality a dislocation lying in the $xy$ plane which is also the slip plane, with normal vector $\Bn = \Be_z$. The Burgers vector is taken in $x$ direction, hence $\Bb = b_x \Be_x$. The dislocation line $\cal C$ is parameterized by line length $s$ as ${\cal C} = \{\Br(s)\}$ and the tangent vector is $\Bt = \partial_s \Br$. The hydrostatic stress of the dislocation then reads (see e.g. \citet{cai2006non})
\begin{equation}
{\rm Tr} \Bsigma(\Br) = - \frac{\mu b(1+\nu)}{4\pi(1-\nu)}\int_{\cal C}  t_y \partial_z \Delta R \ds	
\label{eq:trsigma}
\end{equation}
where $R(s) = |\Br - \Br(s)| = [(x-x(s))^2+(y-y(s))^2+(z-z(s))^2]^{1/2}$, and $\Delta$ is the Laplace operator. The defect-dislocation interaction energy can thus be written as an integral over the dislocation line
\begin{equation}
	E_{\rm s} = \int_{\cal C} \sin\theta(s)e^{\rm s}(s) \ds, 
\end{equation}
where the kernel
\begin{equation}
	e^{\rm s}(s) =  - \frac{v \mu b(1+\nu)}{4\pi(1-\nu)} \partial_z \Delta R(s) = - \frac{v \mu b(1+\nu)}{2\pi(1-\nu)} \frac{(z-z(s))}{R^3}
\end{equation}
describes the interaction between a dilatation center in the origin and the edge component of a dislocation segment located at $\Br$. We note that this kernel is proportional to the $z$ component of the displacement field created by the dilatation center at the segment position: $e^{\rm s}(\Br) = \mu b u_z(\Br)$. Hence, the two problems of calculating the statistical properties of solute-dislocation interaction energies, and of solute displacement fields, are mathematically equivalent. 

\paragraph{Peach-Koehler force and shear stress}

Next we consider the solute-dislocation interaction force, which we derive by using a virtual work argument. During a virtual displacement, the work done by the Peach-Koehler force and the change of the dislocation-solute interaction energy must add up to zero:
\begin{equation}
	\delta E_{\rm s} + \delta W_{\rm PK} = 0. 
	\label{eq:VW}
\end{equation} 
Considering glide motion of the dislocation in its slip plane, the variation $\delta \Br(s)$ of the line shape is carried out within the slip plane only. Moreover, since we are dealing with finite or infinite loops there is no variation of the endpoints of the integral. The work of the Peach-Koehler force can then be written in terms of the solute-induced resolved shear stress $\sigma_{xz}^{\rm s}$ in the dislocation slip system as
\begin{equation}
	\delta W_{\rm PK} =  b \int_{\cal C} \sigma_{xz}^{\rm s}(\Br(s)) \Bg(s).\delta \Br(s) \ds
	\label{eq:WPK}
\end{equation}
where the local glide direction vector is defined as $\Bg = \Bt \times \Bn$ with the components $g_x = -t_y, g_y = t_x$. In order to write the energy variation we note that $\sin \theta = \dy/\ds$. The energy variation is  thus obtained as 
\begin{eqnarray}
	\delta E_{\rm s} &=& \int_{\cal C} \delta[e_{\rm s}(\dy/\ds)] \ds
	= \int_{\cal C} [(\dy/\ds) \delta e^{\rm s} +  e^{\rm s}\delta (\dy/\ds)] \ds\nonumber\\
	&=& \int_{\cal C} [(\dy/\ds) \nabla e^{\rm s} \cdot \delta\Br(s) - {\rm d} e_{\rm s}/\ds] \delta y(s) \ds\nonumber\\
	&=& \int_{\cal C}\nabla e^{\rm s}\cdot[(\dy/\ds)(\delta x(s)\Be_x + \delta y(s) \Be_y)
	- \Bt\delta y(s)]\ds
\end{eqnarray}
where in the second step we have integrated by parts. As $\Bt = (\dx/\ds) \Be_x + (\dy/\ds) \Be_y$ we get
\begin{equation}
	\delta E_{\rm s} = \int_{\cal C} \nabla e^{\rm s}\cdot\Be_x [(\dy/\ds) \delta x - (\dx/\ds) \delta y]\ds = - \int_{\cal C} (\nabla e^{\rm s}\cdot\Be_x) 
	(\Bg(s)\cdot\delta \Br) \ds
\end{equation}
It then follows from Eqs. (\ref{eq:VW}) and (\ref{eq:WPK}) that, since the variation $\delta \Br$ is arbitrary, the resolved shear stress acting on a dislocation segment at $\Br$ is given by
\begin{equation}
	\sigma_{xz}(\Br) = \frac{1}{b} \partial_x e^{\rm s}(\Br)
	= \frac{3 v \mu (1+\nu)}{2\pi(1-\nu)} \frac{xz}{R^5}. 
\end{equation}
This is the shear stress created by a single dilatation center. We now proceed to consider the interaction energy and shear stress field arising from the superposition of infinitely many dilatation centers of random strength and location.

\subsubsection{Interaction between a singular dislocation and many dilatation centers of random strength}

We consider a dilute distribution of defects, modelled as individual dilatation centers with dilatation strengths $v_i$. We assume random independent locations $\Br_i$. In the dilute limit, we can dispense of considerations according to which the defect locations must correspond to substitutional or interstitial sites of an underlying lattice. The energy and shear stress experienced by a dislocation segment at $\Br$ are then given by 
\begin{equation}
	e(\Br) = \frac{\mu b(1+\nu)}{2\pi(1-\nu)} \sum_i v_i \frac{(z-z_i)}{|\Br -\Br_i|^3}\quad,\quad
	\sigma_{xz} = \frac{3 \mu (1+\nu)}{2\pi(1-\nu)} \sum_i v_i \frac{(x-x_i)(z-z_i)}{|\Br -\Br_i|^5}.
\end{equation}
We envisage the dilatation strengths as independent random variables with the statistical properties 
\begin{equation}
\langle v_i \rangle = 0\quad,\quad \langle v_i v_j \rangle =\langle v^2 \rangle \delta_{ij}. 
\end{equation} 
Note that the average dilatation can be always set to zero by subtracting a homogeneous strain field. To evaluate spatial signatures of the fluctating interaction energy and stress fields, it is convenient to move to a continuum representation. To this end, we introduce the singular dilatation strain field
\begin{equation}
	\varepsilon_v(\Br) = \sum_i \frac{v_i}{3} \delta(\Br - \Br_i)
\end{equation}
which allows us to write the fields as
\begin{eqnarray}
	e(\Br) &=&  \int_V {\cal E}(\Br'-\Br) \varepsilon_v(\Br') {\rm d}^3 r'
	\quad,\quad {\cal E}(\Br) = \frac{3\mu(1+\nu)b}{4\pi(1-\nu)} \partial_z \Delta R \label{eq:energykern}\\
	\sigma_{xz}(\Br) &=&  \int_V {\cal S}(\Br'-\Br) \varepsilon_v(\Br') {\rm d}^3 r' \quad,\quad {\cal S}(\Br) = \frac{3\mu(1+\nu)}{4\pi(1-\nu)} \partial_z \partial_x \Delta R 
	\label{eq:stresskern}
\end{eqnarray}
Based on the statistical properties of the field $\varepsilon_v(\Br)$ we can work out the statistical properties of the interaction energy density $e$ and shear stress $\sigma_{xz}$. The correlator of the dilatation strain field 
\begin{equation}
	\varepsilon_v(\Br) = \sum_i \frac{v_i}{3} \delta(\Br - \Br_i)
\end{equation}
is obtained by considering an arbitrary volume $V$ containing $N_V$ atomic sites and evaluating the expectation value of the mean quadratic dilatation strain in $V$, $N_V \langle v^2 \rangle$:
\begin{equation}
	N_V \left\langle v^2 \right\rangle =
	\left \langle \sum_{i, \Br_i \in V} v_i^2 \right \rangle =
	\left \langle\sum_{\stackrel{i,\Br_i \in V}{j,\Br_j \in V}}  v_i v_j \right\rangle
	= 9 \iint_V \langle \varepsilon_v(\Br) \varepsilon_v(\Br') \rangle 
	{\rm d}^3 r {\rm d}^3 r'.
\end{equation}
Since this identity must hold for any volume $V$, we find that the correlation function of the dilatation strain field is given by
\begin{equation}
	\langle \varepsilon_v(\Br)  \varepsilon_v(\Br')\rangle  = \langle v^2 \rangle \frac{\rho}{9} \delta(\Br - \Br')
	\label{eq:straincorr}
\end{equation}
where $\rho$ is the density of defect sites. 

\paragraph{Interaction energy}

Using the correlation function, Eq. (\ref{eq:straincorr}), we can calculate the correlation function of the interaction energy density $e$ as given by Eq. (\ref{eq:energykern}). We obtain 
\begin{eqnarray}
	&&\langle e(\Br')e(\Br'+\Br)\rangle \nonumber\\  &=&\left[\frac{3\mu(1+\nu)b}{4\pi(1-\nu)}\right]^2\iint_{V} \partial_z \Delta R(\Br'' -  \Br')  \partial_{z'} \Delta' R(\Br'''  -\Br'-\Br)\langle \varepsilon_V(\Br'') \varepsilon_v(\Br''')\rangle
	 {\rm d}^3 r''  {\rm d}^3 r''' \nonumber\\
	&=&\left[\frac{\mu(1+\nu)b}{4\pi(1-\nu)}\right]^2 \rho \langle v^2 \rangle\int_{V} \partial_z \Delta R(\Br''-\Br')  \partial_{z'} \Delta' R(\Br''-\Br' - \Br)
	{\rm d}^3 r''  \nonumber\\
	&=& \left[\frac{\mu(1+\nu)b}{4\pi(1-\nu)}\right]^2 \rho \langle v^2 \rangle \Phi_e(\Br)
	\label{eq:energycorr}
\end{eqnarray}
The function $\Phi_e = - (\partial_z \Delta R)*(\partial_z \Delta R)$, where $*$ denotes the convolution operation, can be evaluated in Fourier space using the convolution theorem. With ${\cal F}(R)=-8\pi/k^4$ we obtain ${\cal F}(\Phi_e) = 64 \pi^2 k_z^2/k^4$, hence 
\begin{equation}
	\Phi_e(\Br) = 8 \pi \partial_z^2 R = \frac{8 \pi (x^2+y^2)}{R^3}
\end{equation}
This expression is singular as $\Br \to 0$, as expected because of the singular nature of the single-solute interaction energies. 

\paragraph{Shear stress}

Using Eqs. (\ref{eq:stresskern}) and (\ref{eq:straincorr}), the shear stress correlator derives by a calculation analogous to Eq. (\ref{eq:energycorr}) as
\begin{equation}
	\langle \sigma_{xz}(\Br') \sigma_{xz}(\Br'+\Br)\rangle = \left[\frac{\mu(1+\nu)}{4\pi(1-\nu)}\right]^2 \rho \langle v^2 \rangle \Phi_{\tau}(\Br).
    \label{eq:stresscorr}
\end{equation}
where the stress correlation function is given by
$\Phi_{\tau} = (\partial_z \partial_x \Delta R)*(\partial_z \partial_x \Delta R) = - \partial_x^2 \Phi_e$. In real space we obtain
\begin{equation}
	\Phi_{\tau}(\Br) = 8\pi \left(\frac{1}{R^3}-3\frac{x^2+z^2}{R^5 }+15\frac{x^2z^2}{R^7}\right) 
\end{equation}
while the correlations in the slip plane $z=0$ have the form
\begin{equation}
	\Phi_{\tau}(\Br) = 8\pi \frac{y^2 - 2 x^2}{R^5} 
\end{equation}
As with the energy correlation function, this function is singular in the origin and therefore stands in need of regularization in order to obtain an analytical expression for the mean square shear stress exerted by the solutes in the dislocation slip plane. This will be addressed in the following section.

\subsubsection{Regularized interactions of a dislocation with randomly distributed dilatation centers}
\label{sec:regular}

To regularize the interactions we consider three different approaches. The first two approaches have been proposed to handle the problem of dislocation core singularities in discrete dislocation dynamics, namely (i) core regularization in terms of a continuous distribution of Burgers vector as introduced by \citet{cai2006non}, and (ii) evaluation of core-regularized dislocation stress fields using gradient elasticity of Helmholtz type \citet{po2014singularity,lazar2013fundamentals}. These two approaches, though different in physical motivation, both amount to convoluting the singular dislocation stress field with a regularization function, namely the Burgers vector distribution function $w$ in the core regularization approach of \citet{cai2006non} and the Green's function of the Helmholtz equation in the gradient elasticity approach of \citet{lazar2013fundamentals}. The relationships between both approaches and the classical theory are summarized in Table 1 which is taken with slight modifications from the excellent work of \citet{lazar2013fundamentals} on dislocation gradient elasticity. 
\begin{table}[hb]
	\caption{Structural comparison between different dislocation theories}
	\begin{tabular}{|c|c|c|}
		\hline
		Classical & Core regularized & Helmholtz gradient elasticity \\
		\hline
		$R$ & $R_a$ & $A(R)$ \\
		$\Delta \Delta R = - 8\pi \delta(\Br)$ & $\Delta \Delta R_a = -8\pi w(\Br)$
		& $\Delta \Delta A = - 8\pi G(\Br)$\\
		& $R_a = R*w = \sqrt{R^2 + a^2}$ & $A = R*G = R  + \frac{2a^2}{R}(1-\exp(-R/a))$ \\
		& $w(k)$ no analytical expression & $G(k) = \frac{1}{1 + k^2 a^2}$ \\
		\hline
	\end{tabular}
\end{table}

A third approach by \citet{geslin2021microelasticityI,geslin2021microelasticityII} retains the singular nature of the dislocation stress field and, instead, considers a Gaussian distribution of the dilatation strain around the location of a dilatation center, i.e., the regularization applies to the dilatation strain field. We shall consider this approach for comparison below. 

\paragraph{Core regularization using distributed Burgers vector}

According to the method of \citet{cai2006non}, one considers a dislocation whose Burgers vector is distributed around each point $\Br(s)$ of the dislocation line by a radially symmetric function $\tilde{w}(\Br)$. The ensuing stress field, Eq. (\ref{eq:trsigma}), then contains instead of $R$ the function $R*\tilde{w}$ where * denotes the convolution operation. Going through the same steps as above, we find that 
\begin{equation}
	\Phi_e^{\rm CR} = \partial_z \Delta (R*\tilde{w}) * \partial_z \Delta (R*\tilde{w}) 
\end{equation}
In Fourier space this becomes
\begin{equation}
	{\cal F}(\Phi_e^{\rm CR}) = 64\pi^2 \frac{k_z^2}{k^4} [{\cal F}(\tilde{w})]^2 = 64\pi^2 \frac{k_z^2}{k^4} {\cal F}(w)
\end{equation}
where $w = \tilde{w}*\tilde{w}$. Reverting to real space we find 
\begin{equation}
	\Phi_e^{\rm CR} = 8 \pi \partial_z^2 (R*w).
\end{equation}
According to the proposal of \citet{cai2006non}, the function $w$ (i.e., the Burgers vector distribution convoluted with itself) is adjusted in such a manner that the convolution produces the simple result
\begin{equation}
	R*w(\Br) = \sqrt{R^2 + a^2} =: R_a.
\end{equation}
If we adopt the same procedure, the core regularized energy correlation takes the simple form
\begin{equation}
	\Phi_e^{\rm CR} = 8 \pi \partial_z^2 R_a = 8 \pi \left(\frac{1}{R_a} - \frac{z^2}{R_a^3}\right)
	\label{eq:energycorrCR}
\end{equation}
This expression is evidently regular as $\Br\to 0$. The mean square energy fluctuation derives as
\begin{equation}
	\langle e^2 \rangle = \left[\frac{\mu(1+\nu)}{1-\nu}\right]^2 \rho\langle v^2 \rangle \frac{b^2}{2\pi a} 
\end{equation}

The shear stress correlator is given, in strict analogy to the classical expression, by Eq. (\ref{eq:stresscorr}) where now the Energy correlator is given by Eq. (\ref{eq:energycorrCR}). In real space the stress correlation function thus derives as
\begin{equation}
	\Phi_{\tau}^{\rm CR}(\Br) = 8\pi \left(\frac{1}{R_a^3}-3\frac{x^2+z^2}{R_a^5 }+15\frac{x^2z^2}{R_a^7}\right) 
\end{equation}
which in the slip plane $z=0$ reduces to
\begin{equation}
	\Phi_{\tau}^{\rm CR}(\Br) = 8\pi \frac{a^2 + y^2 - 2x^2}{R_a^5} 
\end{equation}
In cylindrical coordinates $(r,\theta)$, where $\theta$ is the angle with the slip direction ($x$ direction), the correlation function $\Phi_{\tau}$ can be conveniently re-written as
\begin{equation}
	\Phi_{\tau}^{\rm CR}(r,\theta) = \Phi_{\tau}^{\rm CR,L}(r) \cos^2 \theta + \Phi_{\tau}^{\rm CR,T}(r) \sin^2 \theta 
	\label{eq:corrLT}
\end{equation}
where the longitudinal (i.e. in slip direction) and transverse correlation functions are given by
\begin{equation}
	\Phi_{\tau}^{\rm CR,T}(r) = 8\pi\frac{1}{R_a^3}\quad,\quad
	\Phi_{\tau}^{\rm CR,L}(r) = 8\pi\frac{3a^2 - 2R_a^2}{R_a^5}.
\end{equation}
Finally, the mean square shear stress follows by taking the limit $r \to 0$ as
\begin{equation}
	\langle \sigma_{xz}^2\rangle = \left[\frac{\mu(1+\nu)}{(1-\nu)}\right]^2 \frac{\rho\langle v^2 \rangle}{2\pi a^3}. 
\end{equation}

\paragraph{Core regularization using gradient elasticity of Helmholtz type}

To work out the energy density correlation for Helmholtz-type gradient elasticity, we follow the same steps as above. The correlation function then reads
\begin{equation}
	\Phi_e^{\rm HG} = \partial_z \Delta A * \partial_z \Delta A 
\end{equation}
which gives in Fourier space (cf. Table 1)
\begin{equation}
	{\cal F}(\Phi_e^{\rm HG}) = 64\pi^2 \frac{k_z^2}{k^4(1+k^2a^2)^2}  
\end{equation}
The Fourier back transformation of this expression is obtained by decomposition into partial fractions and using Fourier transform pairs given by \cite{lazar2013fundamentals}. We obtain 
\begin{equation}
	\Phi_e^{\rm HG} = 8\pi \partial_z^2 \left[R + \frac{4 a^2}{R}\left(1 -\exp\left(-\frac{R}{a}\right)\right)-a\exp\left(-\frac{R}{a}\right)\right]
\end{equation}
After performing the $z$ derivatives and setting $z=0$ we obtain the energy correlation in the $z=0$ plane as
\begin{equation}
	\Phi_e^{\rm HG} = \frac{8\pi}{R} \left[1 +\exp\left(-\frac{R}{a}\right)+   
\frac{4a}{R}\exp\left(-\frac{R}{a}\right) - \frac{4a^2}{R^2} \left(1-\exp\left(-\frac{R}{a}\right)\right)\right].
\end{equation}
Again, this radially symmetric function is regular as $\lim_{r\to 0}\Phi_e^{\rm HG} = 8\pi/(3a)$. The mean square energy fluctuation derives as 
\begin{equation}
	\langle e^2 \rangle = \left[\frac{\mu(1+\nu)}{1-\nu}\right]^2 \rho \langle v^2 \rangle \frac{b^2}{6\pi a} 
\end{equation}
The shear stress correlation is again obtained as $\Phi_{\tau}^{\rm HG}= - \partial_x^2 \Phi_e^{\rm HG}$. To keep expressions simple, we consider only the correlation within the slip plane $z=0$. After some algebra, the result can be cast into the form of Eq. (\ref{eq:corrLT}), $\Phi_{\tau}^{\rm HG}(r,\theta) = \Phi_{\tau}^{\rm HG,L}(r) \cos^2 \theta + \Phi_{\tau}^{\rm HG,T}(r) \sin^2 \theta$ where the longitudinal and transverse shear stress correlation functions are given by 
\begin{eqnarray}
	\Phi_{\tau}^{\rm HG,L}(r) &=&\frac{8\pi}{r^5}\left[- 2r^2 +24 - \exp\left(-\frac{r}{a}\right)\left(\frac{r^4}{a^4}+6\frac{r^4}{a^4}+22\frac{r^2}{a^2} +48 \frac{r}{a} +48 \right)\right]\nonumber\\
	\Phi_{\tau}^{\rm HG,T}(r) &=& \frac{8\pi}{r^5}\left[r^2 - 12 + \exp\left(-\frac{r}{a}\right)\left(\frac{r^3}{a^3}+5\frac{r^2}{a^2}
	+12 \frac{r}{a} +12 \right)\right].
\end{eqnarray}
from which the mean square shear stress follows  by taking the limit $r\to0$. With $\lim_{r \to 0} \Phi_{\tau} = 8\pi/15$ we obtain
\begin{equation}
	\langle \sigma_{xz}^2\rangle = \left[\frac{\mu(1+\nu)}{(1-\nu)}\right]^2 \frac{\langle v^2 \rangle}{30 \pi V_{\rm at} a^3}. 
\end{equation}

\paragraph{Regularization of the solute dilatation strain}

An alternative approach for regularizing dislocation-solute interactions consists in regularizing the dilatation strain field, Eq. (3), while considering the dislocation as singular. This approach was used by Geslin et. al., \citet{geslin2021microelasticityI,geslin2021microelasticityII} who regularize the dilatation strain field by replacing the $\delta$-distribution of the point dilatation center by a three-dimensional Gaussian function of width $a$, ${\cal G}(r/a) = \exp[-x^2/(2a^2)]/[(2\pi)^{3/2}a^3]$. The solute energy and stress fields are then obtained by convoluting the singular expressions with this Gaussian. 
	
Mathematically, it is irrelevant whether one applies the regularization to the singular dislocation or to the solute - in either case, the interaction functions are convoluted with a regularization function. Thus, the ensuing correlation functions have a structure identical to the ones described above. One finds
\begin{equation}
	\Phi_e^{\rm SR} = \partial_z\Delta(R*{\cal G})*\partial_z\Delta(R*{\cal G})\quad,\quad
	\Phi_{\tau}^{\rm SR} = - \partial_x^2	\Phi_e^{\rm CR}
\end{equation}
We re-phrase the expressions given by \citet{geslin2021microelasticityII} 
in our current terminology. Again, Eq. \ref{eq:corrLT} holds with the longitudinal and transverse stress correlation functions 
\begin{eqnarray}
	\Phi_{\tau}^{\rm SR,L}(r)&=&-\frac{16\sqrt{\pi}}{r^3} \left[\sqrt{\pi}\left(
	1-\frac{12a^2}{r^2}\right){\rm erf}\left(\frac{r}{2a}\right)+\frac{a}{r}
	\left(12 + \frac{r^2}{a^2}\right)r^2 \exp\left(-\frac{r^2}{4a^2} \right)\right].
	\nonumber\\
	\Phi_{\tau}^{\rm SR,T}(r) &=& \frac{8\sqrt{\pi}}{r^3} \left[\sqrt{\pi}\left(
	1-\frac{6a^2}{r^2}\right){\rm erf}\left(\frac{r}{2a^2}\right)+\frac{6a}{r}
	\exp\left(-\frac{r^2}{4a^2}\right)	\right],
\end{eqnarray} 
The mean square shear stress is given by
\begin{equation}
	\langle \sigma_{xz}^2\rangle = \left[\frac{\mu(1+\nu)}{(1-\nu)}\right]^2 \frac{\rho \langle v^2 \rangle}{30 \pi^{3/2} a^3}.
\end{equation}

\paragraph{Comparison}

To compare the different approaches, we note that all correlation functions given above share a common asymptotic behavior, which is the same as for the singular fields and given by Eqs. (\ref{eq:energycorr}) and (\ref{eq:stresscorr}). On the other hand, the values of the correlation functions in the origin, and thereby the magnitude of stress fluctuations, depend both on the value of the regularization length $a$ and on the nature of the regularization function. In practice, the regularization parameter $a$ must be adjusted such as to match the stress fluctuations obtained from atomistic simulation, see  \citet{geslin2021microelasticityI,geslin2021microelasticityII}. For a given level of the mean square shear stress, the different regularization functions therefore lead to different values of the parameter $a$. 

For illustration, we take as reference the value $a_{\rm SR} \approx 1 \text{\AA}$ as determined by \citet{geslin2021microelasticityII} from atomistic simulation. Then, to obtain the same mean square shear stress, the regularization parameters must be chosen as $a_{\rm CR} = 2.98 \text{\AA}$   for core regularization by distributed Burgers vector, and as $a_{\rm GH} = 1.21\text{\AA }$  for core regularization by Helmholtz-type gradient elasticity. 

\begin{figure}
	\centering
	\resizebox{0.8\textwidth}{!}{
		\includegraphics{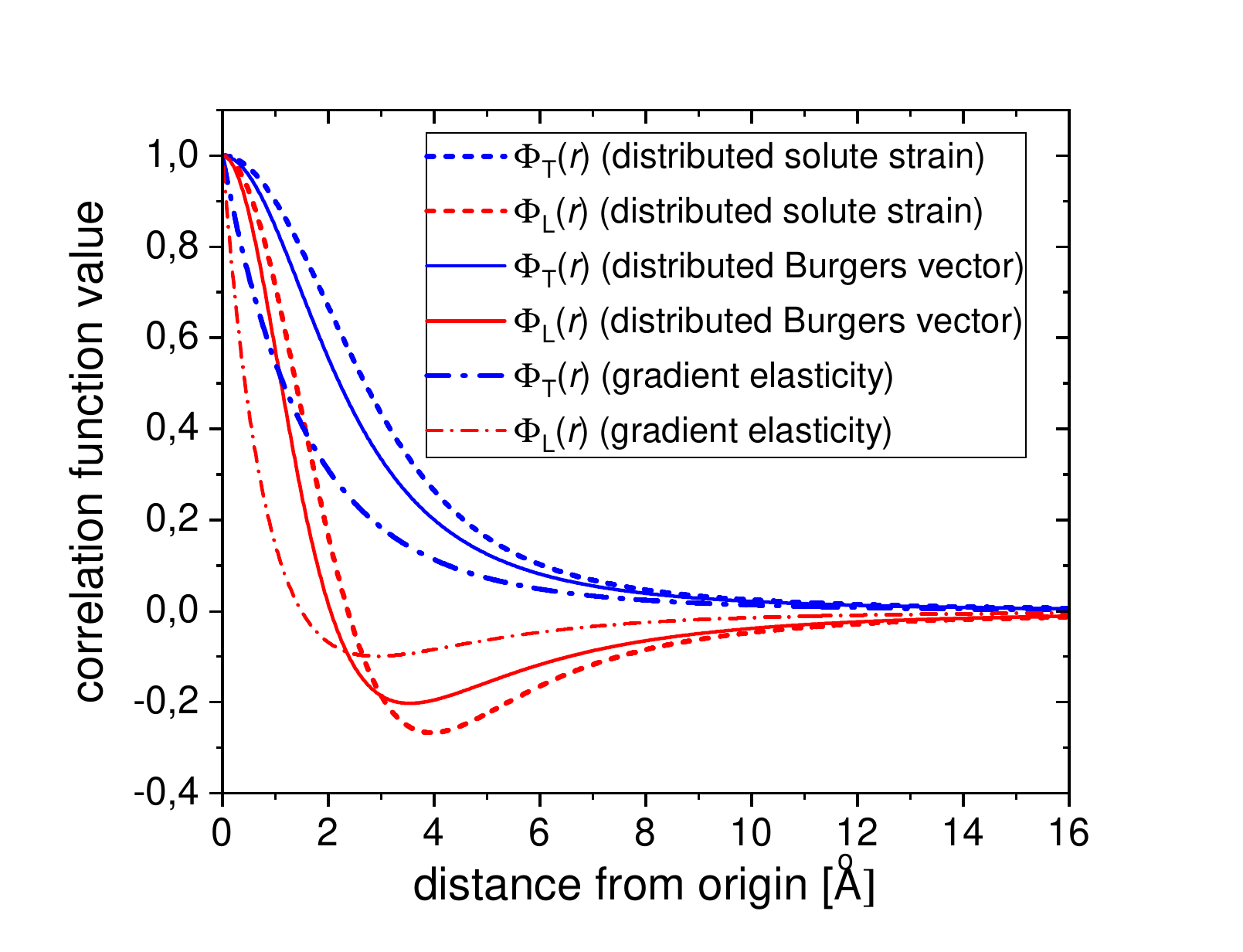}}
	\caption{Longitudinal and transverse correlation functions of the shear stress arising from a random arrangement of solutes as evaluated by different core regularization methods; red: transverse correlation functions, blue: longitudinal correlation functions}
	\label{fig:corrcomp}       
\end{figure}

Correlation functions evaluated with these parameters are compared in Figure \ref{fig:corrcomp}. All correlation functions are normalized so that the value in the origin is equal to 1. With this normalization, the integral under the normalized transverse correlation function defines an important parameter in dislocation pinning theories, namely the transverse correlation length $\xi$ of the fluctuating shear stress field \citep{zaiser2022pinning}. 
This correlation length depends on the manner in which the regularization is performed. For core regularization by distributed Burgers vector we find $\xi^{\rm CR} = 2.84$ \text{\AA}, for regularization by Helmholtz gradient elasticity $\xi^{\rm HG} = 1.83$ \text{\AA},
and for regularization of the solute strain, $\xi^{\rm CS} = 3.25$ \text{\AA}.

In qualitative terms, the different regularization methods lead to broadly equivalent shapes of the correlation functions. In particular, the integral of the longitudinal correlation functions is zero in all cases. We note, however, that the correlation functions evaluated using gradient elasticity have the peculiar feature that they are non-analytic in the origin, where they exhibit a cusped shape. This non-analytic behavior is, however, not a major problem in theories of solute pinning, see the theory of elastic manifold pinning where cusped correlators of the random force acting on an elastic manifold naturally emerge from renormalization theory even when the original correlation function is analytic \citep{nattermann1992dynamics}.

\subsection{Dislocations: Interaction of a dislocation with a random arrangement of dislocations that thread its slip plane
\label{sec:dislocations}}

For the case of random stress fields created by forest dislocations, we consider the shear stress created by an arrangement of rigid straight screw dislocations pertaining to multiple slip systems (we shall assume an fcc lattice structure) and threading the slip plane of the dislocation segment under consideration. While this arrangement seems at first glance highly artificial, we shall later discuss how and to which extent the results carry over to generic arrangements of three-dimensionally curved dislocations. 

\subsubsection{Random arrangement of straight screw dislocations, singular case}

We consider a random arrangement of singular screw dislocations. The dislocations belong with equal probability to one of $N$ Burgers vectors $\Bb_{\alpha}$ and pass with equal and independent probability through any point of the planes with normal vector $\Bs_{\alpha} = \Bb_{\alpha}/b$. Their planar density is denoted as $\rho_{\alpha} = \rho/N$. The vector in the plane with normal $\Bs_{\alpha}$ connecting the origin with the dislocation $(i,\alpha)$ is denoted as $\Br^i_{\alpha}$ with length $R^i_{\alpha}$, and its angle with respect to a fixed reference direction $\Be_{\rm ref,\alpha}$ in that plane is $\phi_{\alpha}^i$. We further define the unit vectors $\Be_r^i = \Br^i_{\alpha}/R^i_{\alpha}$ and $\Be_{\phi}^i = \Bs_{\alpha}\times \Be_r^i$. 

We now consider the resolved shear stress $\tau^{\beta}$ created by this arrangement in the slip system with slip vector $\Bs_{\beta}=\Bb_{\beta}/b$ and normal vector $\Bn_{\beta}$ which, without loss of generality, we take to point in the $x$ and $z$ directions, respectively. Stresses are evaluated without loss of generality in the origin. The stress tensor created by the dislocation located at $\Br^i_{\alpha}$ is 
\begin{equation}
    \Bsigma(\Br^i_{\alpha}) = \frac{\mu b}{2\pi R^i_{\alpha}}
    [\Bs_{\alpha} \otimes \Be_{\phi}^i+
    \Be_{\phi}^i \otimes \Bs_{\alpha}]
\end{equation}
Resolving into slip system $\beta$ by double contraction with the slip system projection tensor gives
\begin{equation}
    \sigma_{xz}(\Br^i_{\alpha}) = \frac{\mu b}{2\pi R^i_{\alpha}}
    [(\Bn_{\beta}\cdot \Bs_{\alpha})(\Bs_{\beta} \cdot \Be_{\phi}^i) + (\Bs_{\beta}\cdot \Bs_{\alpha})(\Bn_{\beta} \cdot \Be_{\phi}^i)] =  \frac{\mu b \eta_{\alpha}\sin(\phi_{\alpha}^i)}{2\pi R^i_{\alpha}} = \mu b \Phi(\Br_{\alpha}).
    \label{eq:sigxzdis}
\end{equation}
where we have chosen the reference direction for dislocations of direction $\Bs_{\alpha}$ as the direction in the slip plane of in slip system $\beta$ in which they create zero resolved shear stress. Values of $\eta_{\alpha}$ for a face-centered cubic lattice and the corresponding reference vectors $\Be_{\alpha,{\rm ref}}$ are compiled in Figure \ref{fig:tetrahedron}.
\begin{table}[htb]
\begin{minipage}{.3\textwidth}
\includegraphics[width=\textwidth]{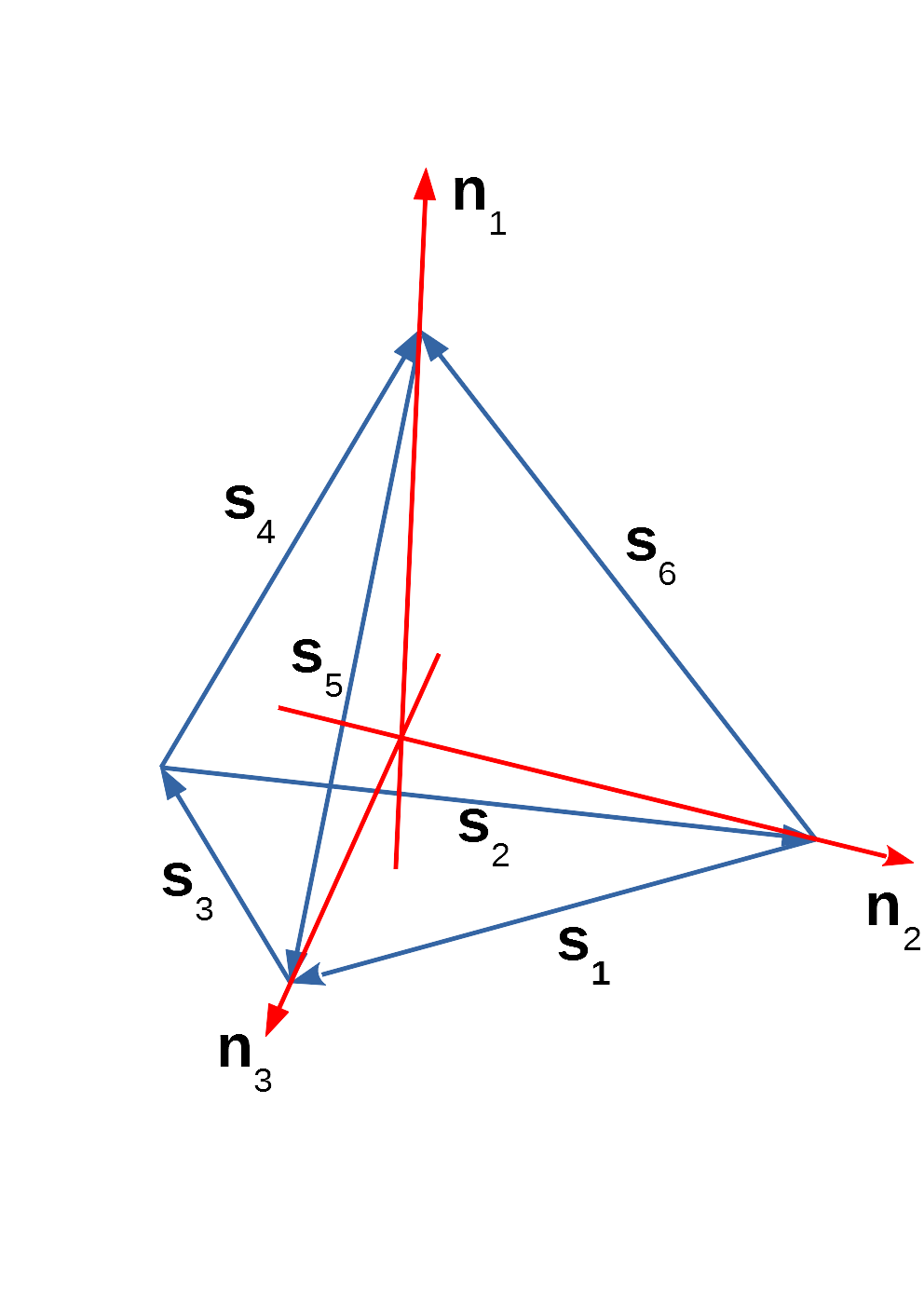}
\end{minipage}
\hfill
\begin{minipage}{.6\textwidth}
	\begin{tabular}{|c|c|c|c|c|c|c|}
    \hline
    slip vector index $\beta$ & 1 & 2 & 3 & 4 & 5 & 6 \\
    \hline
    value of $\eta_i$ 
    & $1$
    & $1/2$ & $1/2$ &  $\sqrt{2/3}$ & $1/2$ & $1/2$\\
    \hline
    &&&&&&\\[-.2cm]
    reference direction 
    & $\Bn_1$ & $\Bn_1$ & $\Bn_1$ & $\Bs_1$
    & $\Bn_2$ & $\Bn_3$ \\[.3cm]
    \hline
    \end{tabular}
\caption{Slip system specific angular parameters for resolved shear 
stress\\created by screw dislocations of orientation $\Bs_i$ in slip system $\Bs_1,\Bn_1$;\\ for nomenclature see figure on the left.}
\end{minipage}
\label{fig:tetrahedron}
\end{table}
To evaluate the spatial correlation function we first note that, for each 'family' $\alpha$ of threading dislocations,  the dislocation arrangement can be characterized by a random uncorrelated pattern of intersection points of density $\rho/N_{\alpha}$ in the perpendicular plane. Correlations between different slip systems are assumed absent. In this case, the stress correlation function in the perpendicular plane is simply obtained by convoluting the single-dislocation stress field, Eq. (\ref{eq:sigxzdis}), with itself. We evaluate this correlation function separately for each 'family' in a coordinate system where $\Bs_{\alpha}$ defines the $z$ axis direction and the reference direction $\Be_{\alpha,{\rm ref}}$ the $x$ axis direction. We introduce the discrete Burgers vector density $b(\Br_{\alpha}) = b\sum_i \delta(\Br_{\alpha}-\Br_{i,\alpha})$  
with the correlation function $\langle b(\Br_{\alpha}) b(\Br'_{\alpha})\rangle = b^2\rho_{\alpha}
\delta(\Br_{\alpha}-\Br'_{\alpha})$. In terms of this function, the stress correlation function for a given 'family' can then be written as
\begin{eqnarray}
C_{\alpha}(\Br_{\alpha}) = \langle \sigma_{xz}(\Br_{\alpha}) \sigma_{xz}(\Br'_{\alpha})\rangle &=& \mu^2 \eta_{\alpha}^2\left\langle \int\int 
b(\Br''_{\alpha})b(\Br'''_{\alpha}) 
\Phi(\Br_{\alpha}-\Br''_{\alpha})\Phi(\Br'_{\alpha}-\Br'''_{\alpha}){\rm d}^2 r''_{\alpha} {\rm d}^2 r'''_{\alpha}\right\rangle\nonumber\\
&=& \mu^2 \eta_{\alpha}^2\int \int \langle b(\Br''_{\alpha})b(\Br'''_{\alpha}) \rangle
\Phi(\Br_{\alpha}-\Br''_{\alpha})\Phi(\Br'_{\alpha}-\Br'''_{\alpha}){\rm d}^2 r''_{\alpha} {\rm d}^2 r'''_{\alpha}\nonumber\\
&=& \mu^2\eta_{\alpha}^2 b^2 \rho
\int
\Phi(\Br_{\alpha}-\Br''_{\alpha})\Phi(\Br'_{\alpha}-\Br''_{\alpha}){\rm d}^2 r''_{\alpha} 
\end{eqnarray}
The last integral is simply a convolution of the function $\Phi$ with itself. In the following we exploit the scaling invariance of dislocation systems \cite{zaiser2014scaling} by measuring all lengths in units of the dislocation spacing, i.e., we introduce non-dimensional space vectors $\tilde{\Br} = \Br_{\alpha} \sqrt{\rho_{\alpha}}$ and conjugate Fourier vectors $\tilde{\Bk} = \Bk_{\alpha}/\sqrt{\rho_{\alpha}}$ where $\Bk_{\alpha}$ is a two-dimensional wavevector in the plane perpendicular to $\Bs_{\alpha}$. 

The correlation function is then evaluated by using the convolution theorem. In Fourier space the scaled Fourier transform of $\Phi$, Eq. \ref{eq:sigxzdis}, is given by
\begin{equation}
  \Phi(\tilde{\Bk}) = \frac{i\tilde{k}_y}{\tilde{k}^2}
\end{equation}
where components $\tilde{k}_y$ is the $y$ component and $\tilde{k}$ the length of $\tilde{\Bk}$. Accordingly the Fourier transform of the correlation function follows as
\begin{equation}
  C_{\alpha}(\Bk_{\alpha}) = \mu^2 \eta_{\alpha}^2 b^2 \rho_{\alpha} \frac{\tilde{k}_y^2}{\tilde{k}^4}
\end{equation}
from which the correlation function itself follows as
\begin{equation}
  C_{\alpha}(\Br_{\alpha}) =
   - \frac{\mu^2 b^2\eta_{\alpha}^2 \rho_{\alpha}}{8\pi} \partial_{\tilde{x}}^2 \left[\tilde{R}_{\alpha}^2 (\gamma_{\rm E} + \ln(\tilde{R}_{\alpha}))\right] 
   = 
  - \frac{\mu^2 b^2\eta_{\alpha}^2 \rho_{\alpha}}{4\pi} \left[\gamma_{\rm E} +\frac{1}{2} + \ln(\tilde{R}_{\alpha}) + \sin^2 \phi_{\alpha} \right]
  \label{eq:corrdis}
\end{equation}
where $\gamma_{\rm E}$ is the Euler-Mascheroni constant. We see that the correlation function has a logarithmic singularity in the origin. Similar behavior was also reported for systems of edge dislocations by \citet{zaiser2002dislocation}. 
The overall correlation function follows by transforming all $C_{\alpha}$ to a common coordinate system and summation over $\alpha$. 

\subsubsection{Random arrangement of straight screw dislocations, core regularized case}

Core regularization of the stress correlation function works analogously to the regularization of the energy correlator for dislocation-solute interactions. We present the explicit calculation for the case of core regularization by Helmholtz gradient elasticity, where the regularization is simply effected by double convolution with the Green's function $G(\tilde{k})$. Thus the Fourier transform of the regularized correlators is
\begin{equation}
  C_{\alpha}(\tilde{\Bk}_{\alpha}) = - \mu^2 \eta_{\alpha}^2 b^2 \rho_{\alpha} \frac{\tilde{k}_y^2}{\tilde{k}^4(1+\tilde{k}^2\tilde{a}^2)^2}
  \label{eq:corrfourierregular}
\end{equation}
Using results of \citet{lazar2013fundamentals}, the real-space correlator can be written as
\begin{equation}
  C_{\alpha}(\tilde{\Br}_{\alpha}) = - \frac{\mu^2 \eta_{\alpha}^2 b^2 \rho_{\alpha}}{8\pi}
  \partial_{\tilde{x}^2} \left[\tilde{R}_{\alpha}^2 (\gamma_{\rm E} + \ln(\tilde{R}_{\alpha}))+8\tilde{a}^2\left(\gamma_{\rm E} + \ln(\tilde{R}_{\alpha}) + K_0\left(\frac{\tilde{R}_{\alpha}}{\tilde{a}}\right)\right)\right]
\end{equation}
Here $K_0$ is a modified Bessel function of the second kind which removes the logarithmic singularity, as an be seen from its expansion for small $x$: $K_0(x)\approx -(\ln x/2 + \gamma_{\rm E})(1+x^2/8) + x^2/4 +{\mathcal O}(x^4)$. At $R_{\rm a}=0$, the function value of the correlation function takes a finite maximum. The value of this maximum defines the mean square resolved shear stress and is simplest obtained by $\tilde{k}$ integration of Eq. (\ref{eq:corrfourierregular}). For a system of size $L$ with periodic boundary conditions, the integration must be constrained to $k>2\pi/{L}$
and we find the result
\begin{equation}
C_{\alpha}(0) = \langle [\sigma_{xz}(\Br_{\alpha})]^2) \rangle = \frac{\mu^2 \eta_{\alpha}^2 b^2 \rho_{\alpha}}{4\pi} \left[\ln\left(\frac{L}{a}\right)-\frac{1}{2}\right]
\end{equation}
Again, the result for multiple slip systems is obtained by summation over $\alpha$.

\subsubsection{Comparison with DDD simulations}

\begin{figure}[htb]
	\centering
	\resizebox{0.8\textwidth}{!}{
		\includegraphics{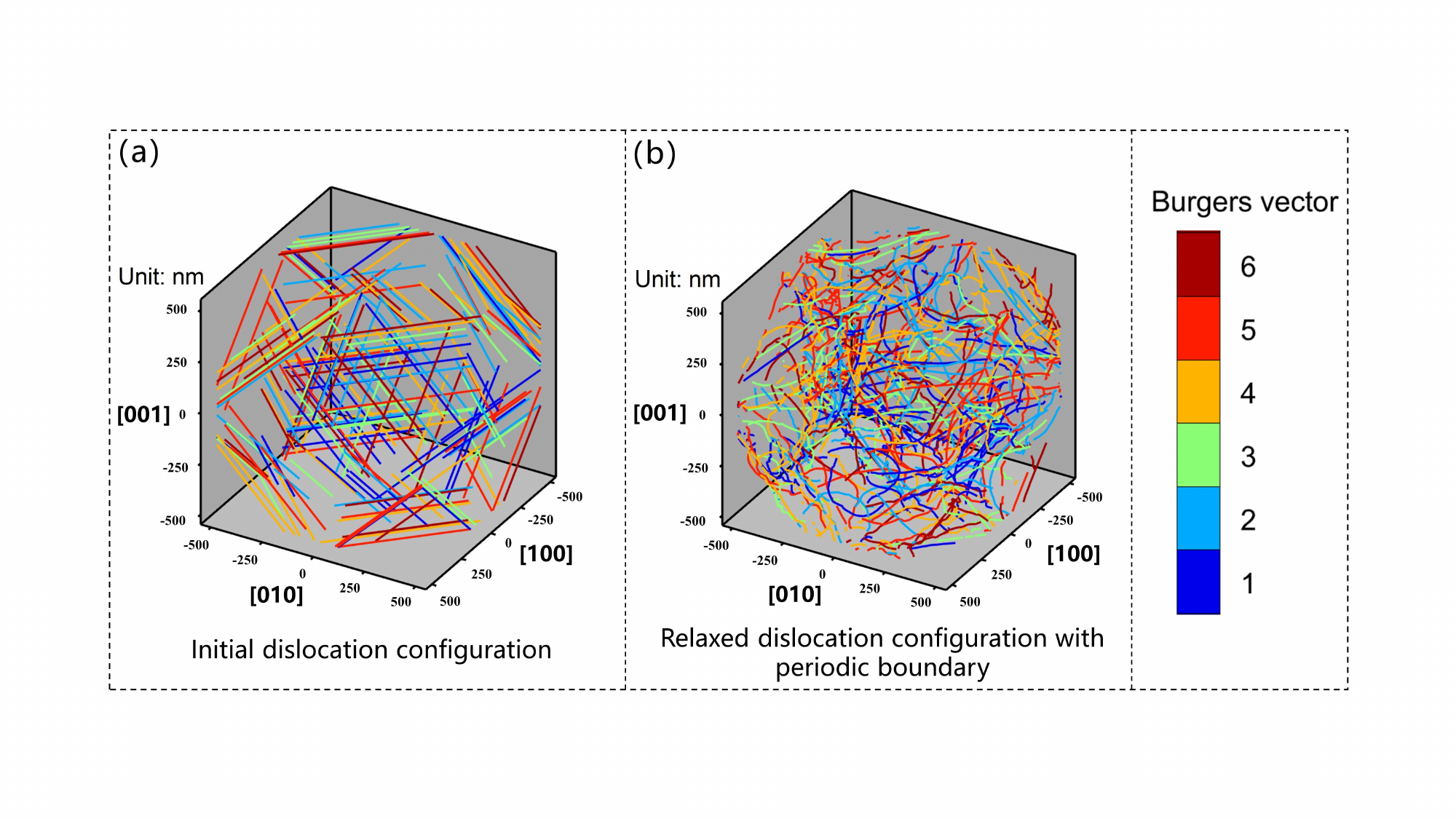}}
	\caption{Simulated dislocation structures; left: initial condition, right: dislocation structure after relaxation}
	\label{fig:dislocations}       
\end{figure}

To assess the relevance of our considerations to generic three-dimensional dislocation systems composed of a network of three-dimensionally connected and in general curved lines, we compare with correlation functions derived from 3D DDD simulations. Simulations were performed using the Multiscale dislocation dynamics plasticity (MDDP) framework \citep{Zbib1998_IJMS,zbib2002multiscale}. A three-dimensional domain of size 1000 nm × 1000 nm × 1000 nm with periodic boundary conditions was simulated. Initially, randomly located pure screw dislocations were evenly distributed over the 4 Burgers vector orientations to produce an overall dislocation density $\rho = 1.5 \times 10^14$ m$^{-2}$. The Burgers vector length was $b=0.25$ nm, the shear modulus $\mu = 48.3$ GPa and Poisson's ratio $\nu = 0.343$. After determining the stress field of this random arrangement, relaxation was performed with periodic boundary conditions until the overall dislocation density and total elastic energy reach a plateau. The stress field was then evaluated for a second time. Initial and relaxed configurations are shown in Figure \ref{fig:dislocations}. 
Stresses were evaluated on a grid mesh of 200 × 200 × 200 nodes. The stress value at each node was calculated by adding up the stresses caused by each dislocation segment, using the equations for core regularized dislocations provided by \citet{cai2006non}. Stresses were evaluated in the initial and in the relaxed states, and resolved shear stresses as well as their spatial correlation functions were determined from the grid data. 

\begin{figure}[htb]
	\centering
	\resizebox{0.95\textwidth}{!}{
	\includegraphics{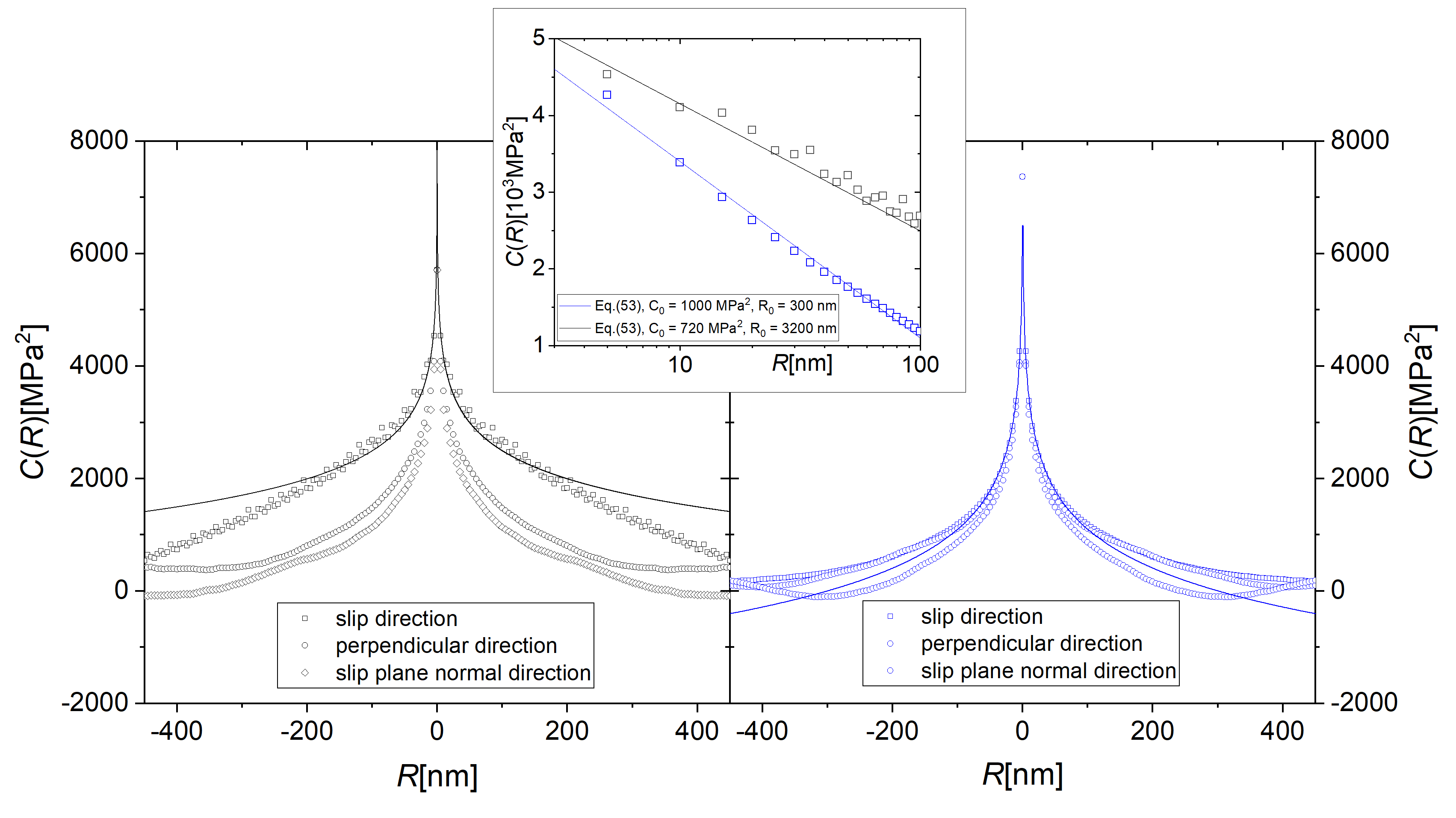}}
	\caption{Correlation functions of dislocation induced resolved shear stress; left: before relaxation, right: after relaxation; inset: semi-logarithmic plot with fits of Eq. (53).}
	\label{fig:stresscorrdis}       
\end{figure}
We first take a look at the data. Figure \ref{fig:stresscorrdis} shows correlations of the resolved shear stress in a representative slip system for the dislocation arrangements shown in Figure \ref{fig:dislocations}, with the left figure representing correlations in the initial state and the right figure in the relaxed state. Correlation functions are shown in three directions, namely the slip direction, the direction perpendicular to the slip direction in the slip plane (edge dislocation direction), and the slip plane normal direction. 
We make the following observations:
\begin{itemize}
\item 
The correlation function before relaxation is strongly anisotropic. Stress correlations in slip direction span the entire system. This is a direct effect of the initial condition (straight system-spanning screw dislocations) since the line direction of each didlocation coincides with the slip direction of two slip systems.
\item
After relaxation, correlations become much more isotropic. This is consistent with the much more isotropic character of the dislocation arrangement which results as dislocations acquire curvature, rotate into non-screw orientations, and form junctions (Figure \ref{fig:dislocations}, right). 
\item
During relaxation, the characteristic range of stress correlations decreases, consistent with the idea that relaxation leads to screening of long range stresses. 
\item
The long-distance behavior of the correlations is controlled by boundary conditions and thus not described by the theoretical expression for bulk behavior.
\end{itemize}
To quantitatively analyze the results shown in Figure \ref{fig:stresscorrdis}, we first observe that, with a grid resolution of 5nm for the stress pattern, it is impossible to observe the effects of core regularization, as the core radius is one order of magnitude smaller than the grid resolution. We therefore base our analysis on the singular expression for the correlator, Eq. (\ref{eq:corrdis}). Second, we observe that, by pulling the angle dependent and constant terms into the logarithm, and summing over all Burgers vectors, the correlation function can be written in the form
\begin{equation}
  C(\Br) = - C_0 \ln\left(\frac{R}{R_{0}(\theta,\phi)}\right) 
  \quad\quad
   C_0 = - \frac{\mu^2 b^2\langle \eta_{\alpha}^2 \rangle \rho}{4\pi}. 
  \label{eq:corrdis1}
\end{equation}
Fitting this expression to the central part of the correlation function ($R < 100$nm) provides a good match, demonstrating the logarithmic nature of the correlations. Before relaxation, we find from the fit a pre-factor $C_0 \approx 720$MPa$^2$, in good agreement with the value of $C_0 = 771$MPa$^2$ deduced from Eq. (53). After relaxation $C_0 \approx 990$MPa$^2$, which is consistent with the observed increase in dislocation density as initially straight disloations bow into curved shapes. At the same time, the fitted range parameter $R_0$ decreases by about one order of magnitude from $R_0 \approx 3000$nm before relaxation to $R_0 \approx 300$ nm after relaxation. This is consistent with the idea that the correlation range before relaxation is controlled by the system size, whereas after relaxation, it reduces to about 3 mean dislocation spacings, i.e., a typical range for dislocation-dislocation correlations. 

\section{Probability densities of random elastic fields}\label{sec:3}

Besides correlation functions, another set of important statistical signatures of the elastic fields generated by randomly arranged defects are the probability densities (or equivalently the distribution functions) of local field variables such as displacement, stress/strain, and dislocation segment energy. With respect to displacement and stress/strain fields of randomly arranged dilatation centers, \citet{geslin2021microelasticityI} make the assertion that these fields are Gaussian distributed. As we shall see, this assertion is not fully correct. 

Evaluating the statistics of random fields arising from superposition of the fields of infinitely many stochastic point sources is a problem that arises in many areas of science. We mention the stress field of a random arrangement of straight parallel dislocations \citep{groma1998probability,beato2005statistical}, the velocity field in a random arrangement of vortices \citep{kuvshinov2000holtsmark}, the gravitational force field in a random universe \citep{chandrasekhar1943stochastic} and the electrical field of a random arrangement of ions \citep{holtsmark1919verbreiterung}. Here we use the mathematical methods developed by \citet {holtsmark1919verbreiterung} and \citet{chandrasekhar1943stochastic} to deal with the problems of displacement, energy and stress statistics in a random solid solution, and with the problem of stresses in a random dislocation arrangement. 

We use a generic formulation for a variable ${\cal H} \in \{e,u_z,\sigma_{xz}\}$ created by superposition of the fields of discrete defects that are randomly located in a $d$ dimensional space ($d =3$ for dilatation centers, $d=2$ for straight forest dislocations). The defect strength $s$ is considered a (discrete or continuous) random variable. We first consider the case of singular (non-regularized) defect fields, as first studied in the context of electric fields in a plasma by Holtsmark \cite{holtsmark1919verbreiterung}. We shall denote this type of stochastic fields, which serve as a reference for the corresponding regularized fields, as Holtsmark fields. 

\subsection{Singular Holtsmark fields}

Without loss of generality we consider the value of the variable ${\cal H}$ in the origin. Using spherical coordinates, this can be written as
\begin{equation}
	{\cal H} = \sum_i s_{{\cal H},i} \frac{\eta_{\cal H}(\theta_i,\phi_i)}{R_i^{d_{\cal H}}} 
	\label{eq:holtsmarkfield}
\end{equation}
Here, $s^{\cal H}_i$ are scalar source strengths of source $i$ for the field ${\cal H}$, the functions $\eta_{\cal H}$ express the dependence of the source field on the angular coordinates $(\theta_i,\phi_i)$ of the source, $R_i$ is the distance of the source from our reference point in the origin, and the exponent $d_{\cal H}$ characterizes the radial decay of the source fields. For the fields under consideration, the respective values of these functions and parameters can be obtained from Eqs. (\ref{eq:energykern}) and(\ref{eq:stresskern}). The results are compiled in Table 2. The sources are assumed to be located randomly with density $\rho$.

\begin{table}[hb]
	\caption{Parameters for the evaluation of different Holtsmark fields generated by a random arrangement of defects}
	\begin{tabular}{|c||c|c|c|c|}
		\hline
		Field ${\cal H}$ & Displacement $u_z$ & Segment energy $e$ & Shear stress $\sigma_{xz}$ & Shear stress $\sigma_{xz}$\\
        & (Dilatation Centers) & (Dilatation Centers) & (Dilatation Centers) & 
        (Forest dislocations)\\
		\hline\hline
		Source strength $s_{{\cal H},i}$ & 
		$s_{u,i} = - v_i \frac{\mu b(1+\nu)}{2\pi(1-\nu)}$&
		$s_{e,i}= - v_i \frac{(1+\nu)}{2\pi(1-\nu)}$ & 
		$s_{\sigma_{\rm s},i} = v_i \frac{3 \mu (1+\nu)}{2\pi(1-\nu)} $ & $s_{\sigma_{\rm d},i} =  \frac{\mu b_i}{2\pi}$ \\
		\hline
		Angular function $\eta_{{\cal H}}$ & $\eta_{u} = \cos \theta_i$
		& $\eta_{e} = \cos \theta_i$ & $\eta_{\sigma} = \cos \theta_i \sin \theta_i \cos\phi_i$ & $\cos \theta_i$ \\
		\hline
        Space dimensionality $d$ & $d=3$ & $d=3$ & $d=3$ & $d=2$\\
        \hline
		Exponent $d_{\cal H}$ & $d_u = 2$ & $d_e = 2$ & $d_{\sigma}=3$ & $d_{\sigma}=3$\\
		\hline
	\end{tabular}
\end{table}
Probability distributions of the Holtsmark fields can be evaluated using a variation of Markow's method which we describe in detail in Appendix A. Here we give only the main results. The distribution is evaluated in terms of its generating function $F(Q)$ as
\begin{equation}
	p(\tilde{\cal H}) = \frac{1}{2\pi} \int \exp(i\tilde{Q}\tilde{\cal H}) F(\tilde{Q}) {\rm d}\tilde{Q}
\end{equation}
where $\tilde{\cal H} = {\cal H}/H_0, \tilde{Q} = Q H_0 $. Expressions for the generating functions $F$ and the characteristic fields $H_0$ are evaluated in the Appendix; the characteristic fields $H_0$ are in all cases of the order of the mean field exerted by a source that is at the characteristic distance $\rho^{-1/d}$, it can thus be considered a measure of the nearest-neighbor interaction. 

Generally speaking, the Holtsmark distributions considered here are symmetrical, Levy $\alpha$-stable distributions where $\alpha$ depends on the decay exponent of the source field and on the space dimension. The distributions exhibit power-law decay in the large-field tails, $p({\cal H}) \propto {\cal H}^{- \beta}$. $\alpha$ and the tail exponent are related to $d_{\cal H}$ and space dimensionality $d$ via
\begin{equation}
    \alpha = \frac{d}{d_{\cal H}}\quad,\quad
    \beta = 1 + \alpha .    
\end{equation} 
Accordingly, the second moments of the distributions diverge in all considered cases. Relevant expressions are compiled in Table 3. 
\begin{table}[hb]
	\caption{Characteristic field strengths, generating functions and probability densities for Holtsmark fields generated by a random arrangement of dilatation centers. For the cases with $\alpha_{\cal H}=3/2$ marked with an asterisk *, the probability density can be expressed in terms of hypergeometric functions but the result is lengthy and unprofitable to use.} 
	\begin{tabular}{|c||c|c|c|c|}
		\hline
		Field ${\cal H}$ & Displacement $u_z$ & Segment energy $e$ & Shear stress $\sigma_{xz}$ & Shear stress $\sigma_{xz}$\\ 
        & (Dilatation Centers) & (Dilatation Centers) & (Dilatation Centers) & 
        (Forest dislocations)\\
		\hline\hline
		&&&&\\
		\\[-3ex]
		Characteristic field $H_0$
		 & $\displaystyle2\pi \left(\frac{4 \langle |s_u|^{3/2} \rangle \rho }{15}\right)^{2/3}$
		&
	    $\displaystyle2\pi \left(\frac{4 \langle |s_e|^{3/2} \rangle \rho }{15}\right)^{2/3}$ & 
		$\displaystyle\frac{4 \pi \langle |s_{\sigma_{\rm s}}| \rangle \rho }{9}$ & $\displaystyle_{\sigma_{\rm d}} \left(\frac{\pi \langle\eta_{\alpha}^2\rangle}{2} \rho\right)^{1/2}$
         \\[10pt]
		\hline
		&&&&\\
		\\[-3ex]
		Generating function $F(\tilde{Q})$ 
		 & $\exp(-|\tilde{Q}|^{3/2})$
		& $\exp(-|\tilde{Q}|^{3/2})$ & $\exp(-|\tilde{Q}|)$ & $\exp(-|\tilde{Q}|^2 \ln(\chi/\tilde{Q})$\\[10pt]
		\hline
		&&&&\\
		\\[-3ex]
		Probability density $p(\tilde{\cal H})$ & * & * & $\displaystyle \frac{2}{\pi(1+\tilde{\cal H}^2)}$ & *    \\[10pt]
		\hline
        Large ${\cal H}$ asymptotics & 
        $\propto 1/{\cal H}^{5/2}$ & $\propto 1/{\cal H}^{5/2}$
        & $\propto 1/{\cal H}^{2}$  & $\propto 1/{\cal H}^{3}$ \\
        \hline
	\end{tabular}
\end{table}

\subsection{Core regularized Holtsmark fields}

The different regularization strategies discussed in Section \ref{sec:regular} amount to convoluting the singular point defect fields with a regularization function ${\cal G}(r/a)$ which, depending on the regularization method, has different physical meanings -- a Burgers vector distribution, a Gaussian distribution of eigenstrains around the dilatation/compression center, or the Green's function of the Helmholtz equation. In all cases, the regularization function has a characteristic range $a$. The regularized field relates to the singular Holtsmark field by
\begin{equation}
	\tilde{\cal H} = {\cal H}*{\cal G} = \sum_i s_{{\cal H},i} \frac{g_{\cal H}(R_i/a)\eta_{\cal H}(\theta_i,\phi_i)}{R_i^{d_{\cal H}}} 
	\label{eq:holtsmarkfieldreg}
\end{equation}
where the core functions $g_{\cal H}(R/a) = {\cal G}(R/a)*(1/R^{d_{\cal H}})$ remove the singularity of the Holtzmark source fields at the source location. These functions depend on the regularization function ${\cal G}$ and on the radial decay exponent of the singular source field. 

For core regularized fields, no analytical expressions of the generating function or the associated probability density can be given. However, some useful series expansions of the generating function can be derived as discussed in Appendix B2 which govern the behavior in the limits of small and large values of the Fourier variable $Q$. One writes the generating function as $F_{\cal H}(Q) = \exp[- \rho C_{\cal H}(Q)]$ and seeks asymptotic expansions for the function $C_{\cal H}(Q)$. 
In the limit of small $Q$ the function $C_{\cal H}$ can be represented in terms of its Taylor expansion around $Q=0$ as
\begin{equation}
	C_{{\cal H}}(Q) = \sum_{k=1}^{\infty} \frac{(-1)^{k+1}}{(2k)!}\frac{ \langle v^{2k} \rangle}{a^{2kd-3}} Q^{2k} \int_\Omega\int_0^{\infty}  \left(\frac{|\eta| g_{{\cal H}}(U)}{U^{d}}\right)^{2k} U^2 {\rm d}U {\rm d}\Omega.
\end{equation}
where $U = R/a$. The limit of the singular Holtsmark fields is recovered by setting $g = 1$, however, in this limit all terms of the Taylor expansion diverge.

Eq. () allows to directly evaluate the second moment of the distribution, 
\begin{equation}
	\langle {\cal H}^2 \rangle = -\partial_{Q}^2 A_{{\cal H}}(Q)|_{Q=0} = - \rho 
	\partial_{Q}^2 C_{\cal H}(q)|_{q=0} = \frac{\rho \langle v^{2} \rangle}{a^{2d-3}} \int_\Omega\int_0^{\infty}  \left(\frac{|\eta| g_{\cal H}(U)}{U^{d-1}}\right)^{2} {\rm d}U {\rm d}\Omega.
\end{equation}

Results for the different fields are compiled in Table 4, for the case of core regularization using distributed dilatation strain. It must be emphasized that the existence of the second moment of the core regularized Holtsmark fields -- which is controlled by the small-$Q$ asymptotics of the generating function -- does not guarantee a Gaussian shape of the corresponding probability distribution. This can be seen when we consider the asymptotic behavior of the generating function for {\em large} $Q$. For a Gaussian probability distribution, the generating function has equally Gaussian shape. For the core-regularized Holtsmark fields, on the other hand, it is shown in Appendix B that the large-$Q$ asymptotics is identical with that of the corresponding non regularized fields, indicating non-Gaussian behavior. 

\subsection{Comparison with simulations}

\subsubsection{Shear stress field of solutes}
\begin{figure}[htb]
	\centering
	\resizebox{0.8\textwidth}{!}{
		\includegraphics{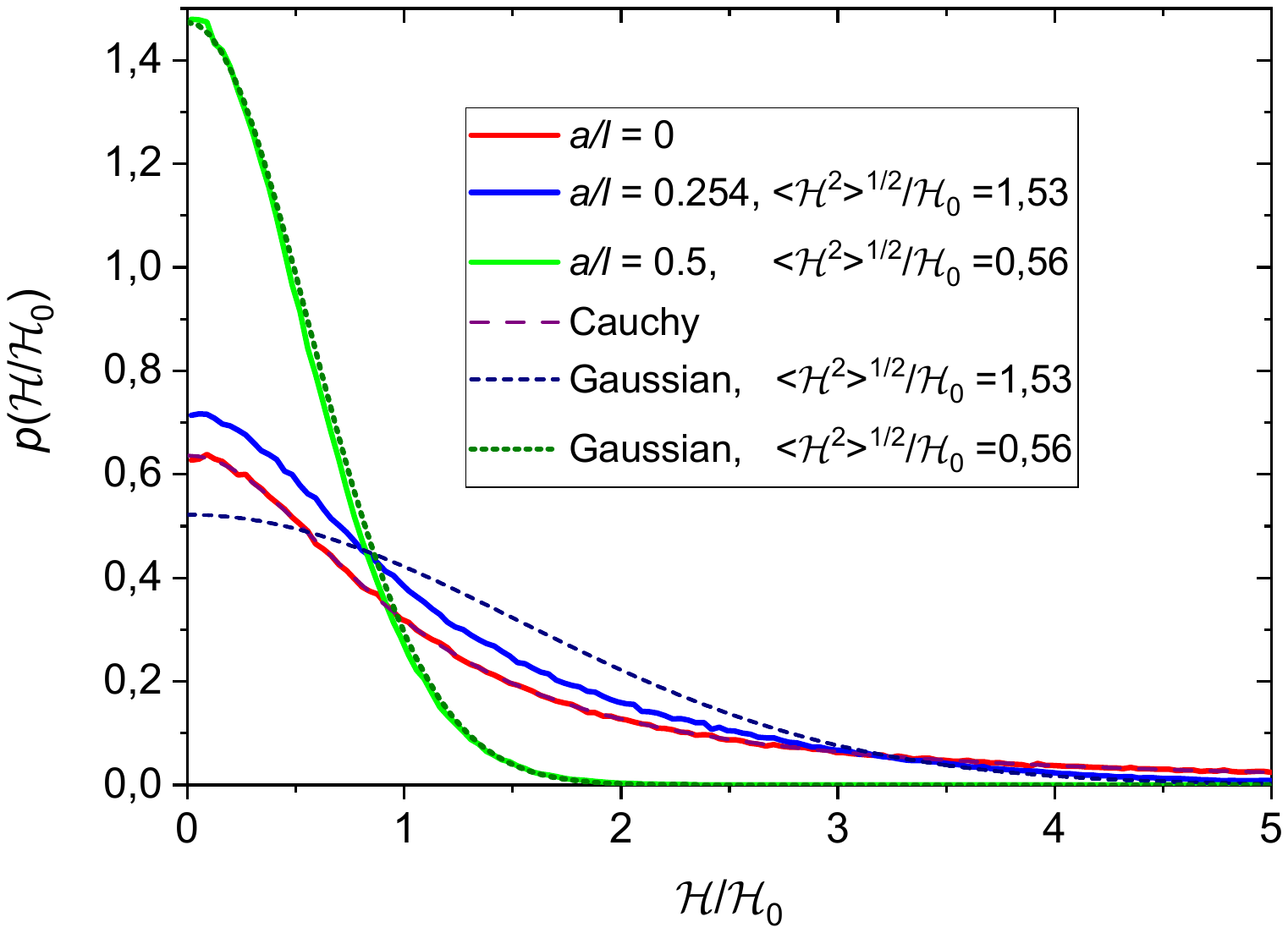}}
	\caption{Probability distributions (probability density funtions) for the resolved shear stress arising from a random arrangement of dilatation centers of density $\rho = 1/l^3$,  dilatation strains are regularized using a Gaussian regularization function of width $a$, for different values of the parameter $a/l$. For comparison, a Cauchy distribution and Gaussian distributions of the same variance as the core-regularized distributions are shown}
	\label{fig:Holtsmark}       
\end{figure}
For shear stress fields created by randomly distributed solutes, the behavior of the core-regularized distributions is governed by the ratio between two characteristic lengths, or equivalently, two characteristic fields. The first length is the source spacing $\rho^{-1/3}$, which governs the characteristic nearest-neighbor field ${\cal H}_0$, whereas the second length is the regularization length $a$, which controls the root-mean-square field $\langle{\cal H}^2\rangle^{1/2}$. Accordingly, we can distinguish two limit cases:
(i) In the limit where the regularization length is much larger than the solute spacing, $\rho^{1/3} a \gg 1$ (${\cal H}_0^2 \ll \langle{\cal H}^2\rangle$), the behavior approaches that of a Gaussian distribution; (ii) in the opposite limit $\rho^{1/3} a \ll 1$ (${\cal H}_0^2 \gg \langle{\cal H}^2\rangle$), the behavior is similar to that of a Holtsmark distribution which is truncated in its outermost tail. For illustration, probability distributions of regularized Holtsmark-like stress fields with Gaussian regularization function and different values of $\rho^{1/3} a$ as determined by Monte Carlo sampling are shown in \ref{fig:Holtsmark}. The Holtsmark-like fields were created by distributing $8 \times 10^3$ stress sources of density $\rho = 1$ and strength $s \pm 1$ randomly over a cube of edge length $20$ with periodic boundary conditions and evaluating the field in the origin for $10^6$ realizations. It can be seen that the non regularized Holtsmark field $(a = 0)$ perfectly follows the theoretical prediction of a Cauchy distribution (Table 4). The probability distribution of the regularized field with $\rho^{1/3} a = 0.5$ is well represented by a Gaussian distribution of the same variance. The field with $\rho^{1/3} a = 0.254$, which is the value for which the correlation functions in \ref{fig:corrcomp} have been evaluated, lies in between. This distribution is best described as a tail-truncated Cauchy distribution. While its second moment is finite, the central part is not well described by a Gaussian shape.

\subsubsection{Shear stress field created by forest dislocations}

\begin{figure}[thb]
	\centering
	\resizebox{0.95\textwidth}{!}{
	\includegraphics{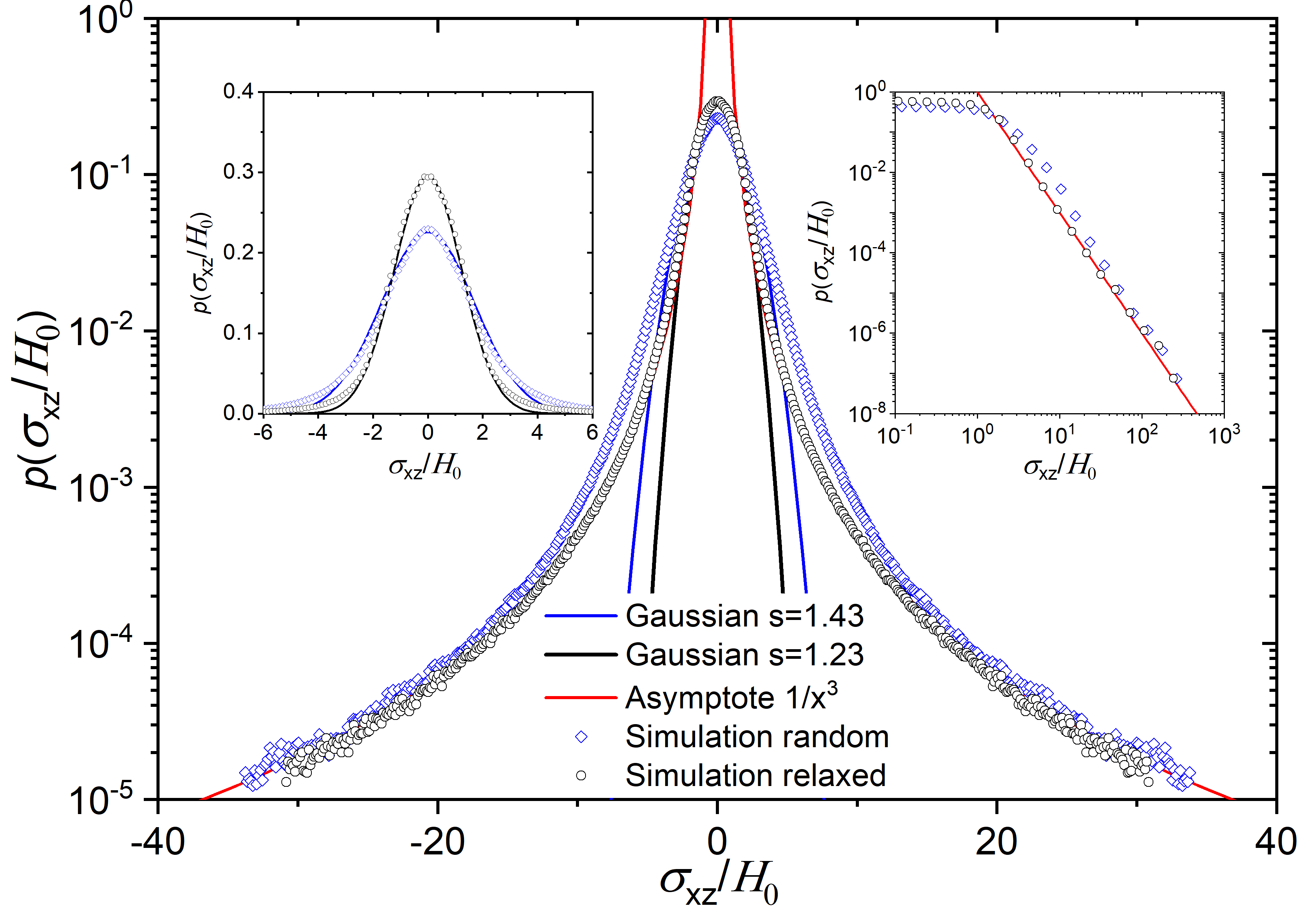}}
	\caption{Probability distributions (stress histograms) for the resolved shear stress arising from random arrangements of dislocations as described in Section 2.2.3, all data refer to shear stresses resolved in the [1-11]/[110] slip system; black data points: streses of randomly arranged straight screw dislocations before relaxation, blue data points: stresses after relaxation at zero external load; left inset and black and blue curves: Gaussian functions with variance $s$ fitted to the central part of the probability distributions; right inset and red curve: analytical asymptote given in Table 4 for large stress fields; the stress parameter $H_0$ is given in Table 4. 
    }
	\label{fig:HoltsmarkDislocations}       
\end{figure}

Probability density functions for the shear stress field created on a slip system by randomly distributed (forest) dislocations were determined empirically from 3D DDD simulations as described in Section 2.2.3. A system of randomly distributed straight screw dislocations was considered alongside the same system after relaxation at zero applied load. Results are shown in Figure\ref{fig:HoltsmarkDislocations}.

The distributions can be best described as composed of a Gaussian central part and a power-law tail which defines the asymptotic behavior for large stresses. The power-law tail where $p(\sigma_{xz}) \propto 1/\sigma_{xz}^3$, constitutes a signature of Holtsmark behavior. This tail is in fact universal and does {\em not} depend on the presence or absence of correlations in the dislocation arrangement, as can be seen in the right inset of Figure \ref{fig:HoltsmarkDislocations}. The central quasi-Gaussian part, by contrast, depends on the mutual arrangement of dislocations. In relaxed dislocation arrangements the width of this part is controlled by the correlation length of the dislocation network and thus by the characteristic dislocation spacing. In random unrelaxed dislocation structures, the width of the central part of the distribution scales in proportion with the logarithm of system size.

\section{Discussion and Conclusions}

Our investigation has demonstrated the complex nature of the statistics of defect induced internal stress fields, which considerably deviates from the assumption of short-range correlated Gaussian random fields as often assumed in the literature. 

What we have investigated in this study are the spatial correlations and the statistics of space-dependent fields created by randomly distributed defects, in particular of the shear stress fields acting in the slip systems of dislocations. In this investigation, the defect arrangement is considered static and an evident question concerns the relevance of our results for the {\em dynamics} of dislocations. 

In the context of dislocation-solute interactions, this question has been investigated in depth in theoretical studies of solute pinning, where the solute-induced forces are envisaged in terms of static "quenched" disorder. The effect of extended spatial correlations of the shear stress was in this context studied by  \citet{zaiser2022pinning} and recently in detail by \citet{rodney2024does}. These studies explicitly or implicitly assume Gaussian statistics of the random shear stress field. It remains to be investigated whether the present findings of fat tails in the probability distribution of solute-induced resolved shear stresses, which become particularly prominent in the limit of low solute concentrations, require to make adjustments to theories of solute pinning. 

\begin{figure}[b]
	\centering
	\resizebox{0.95\textwidth}{!}{
	\includegraphics{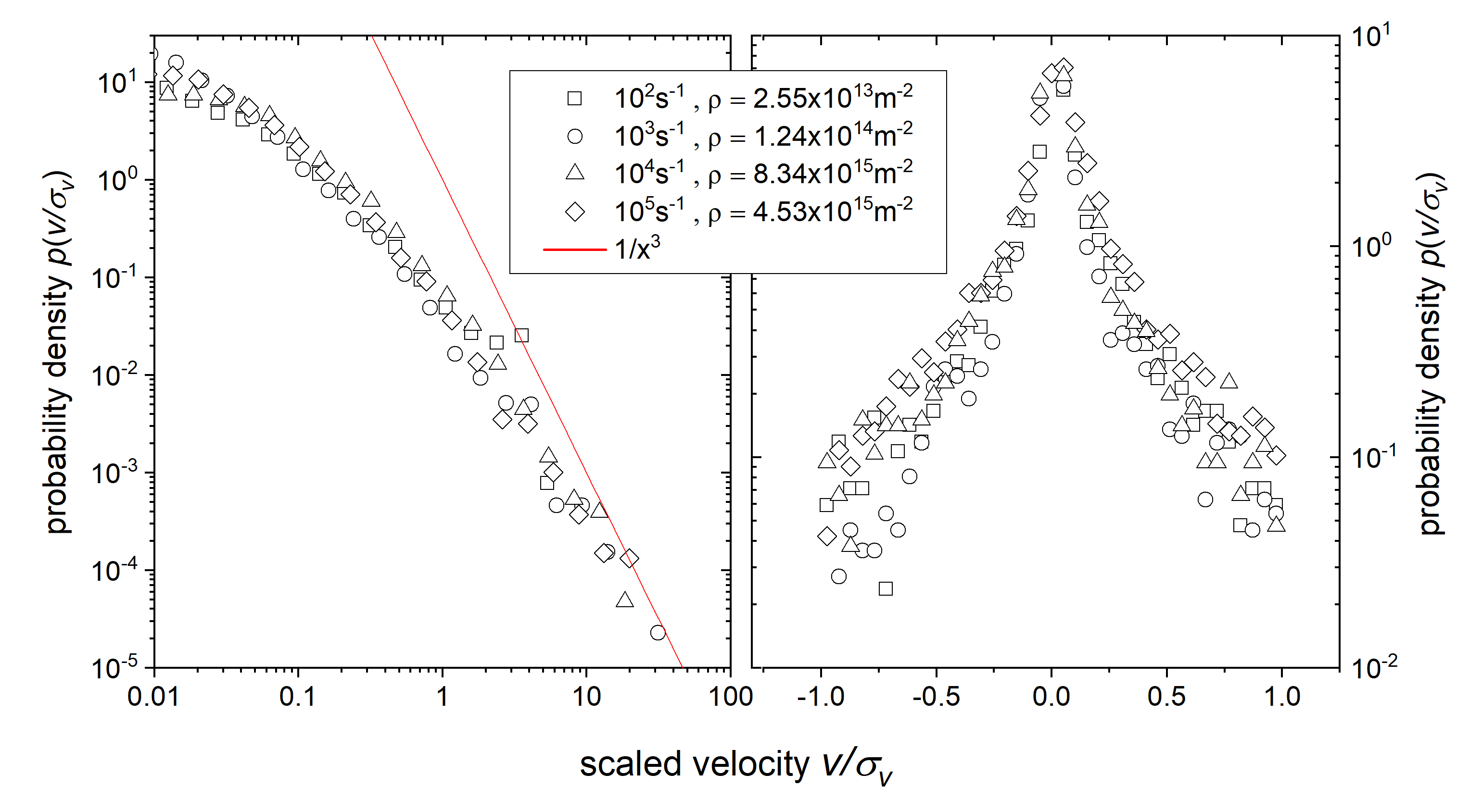}}
	\caption{Probability distributions of dislocation velocities determined for different dislocation densities and imposed strain rates as shown in the legend; all distributions have been normalized by the velocity variance; left: absolute values of the glide velocity in double-logarithmi representation, right: sign dependent velocity distribution (segments moving opposite to the imposed strain rate are shown with negative velocities), data after \citet{fan2021strain}.}
	\label{fig:DislocationVelocity}       
\end{figure}
Even more interesting is the question of dynamic phenomena in the context of dislocation induced stress fields. If we assume overdamped dislocation motion controlled by phonon drag, the local resolved shear stress acting on a dislocation segment is directly proportional to its glide velocity. However, what we have evaluated here is the statistics of resolved shear stresses at fixed, randomly chosen points in space, whereas evaluating the statistics of dislocation segment velocities is equivalent to evaluating the spatial statistics of shear stresses {\em conditional on these spatial points being occupied by dislocation segments}. Moving from the statistics of dislocation induced resolved shear stresses to the statistics of dislocation velocities thus requires a theory of dynamic correlations in dislocation systems, which is beyond the scope of the present study. Nevertheless, simulation results indiate a fairly straightforward correlation between both types of statistics. \citet{fan2021strain} performed simulations of dislocation dynamics over a very wide range of strain rates and dislocation densities and found two distinct dynamic regimes. In a rate controlled regime at high strain rates and/or low dislocation densities, dislocation motion is controlled by high external stresses required to move dislocations at sufficiently high velocities to accomodate the high imposed strain rates. In this regime, internal stresses and dislocation interactions are not important. In an internal stress controlled regime at low strain rates and/or high dislocation densities, on the other hand, dislocation motion is mostly controlled by dislocation interactions. Dislocation velocity distributions determined in this regime are shown in Figure \ref{fig:DislocationVelocity} for different strain rates and dislocation densities. After normalizing all velocities by their rate and dislocation density dependent variance, one observes a universal shape of the probability distributions which shows two remarkable features. 

First, we find that the mean velocity required to accommodate the imposed strain rate is much smaller than the variance of the instantaneous dislocation velocities. In fact the distribution is nearly symmetric (Figure \ref{fig:DislocationVelocity}, right) , which implies that at any moment almost half of the dislocation segments move, under the influence of their mutual interactions, in directions opposite to that imposed by the external stress alone. It was shown by \citet{fan2021strain} that this unexpected type of fluctuating motion is in fact needed to ensure that the macroscopically averaged dissipation matches the microscopic power dissipated by the motion of the dislocation lines.  

Second, the distribution exhibits a power law tail at high velocities (Figure \ref{fig:DislocationVelocity}, left) with an exponent that matches the tail exponent of the internal stress distribution. It is thus very likely that the Holtsmark form of this distribution has a direct impact on the statistics of dislocation motion even under external driving. More work is needed to establish the precise nature of the dynamic correlations emerging under Holtsmark distributed mutual interactions of disloccations. 

\section*{Acknowledgement}

The authors acknowledge support by the German-China Research Office under Exchange Grant No. M-0511. During the initial parts of this investigation M.Z. was supported by DFG under Grant No. Za 171 11-1. R.W. also acknowledges support by  National Natural Science Foundation of China (Grant No. 12272299).

 \newpage

\appendix
\setcounter{equation}{0}
\renewcommand{\theequation}{\Alph{section}.\arabic{equation}}

\section{Probability distributions of singular defect fields - Method of calculation}

For evaluating the statistics of solute displacements or interaction energies we adopt the procedure of Chandrasekhar \cite{chandrasekhar1943stochastic} and of Holtsmark \cite{holtsmark1919verbreiterung}. In a first step we consider singular fields (point-like solutes interacting with line-like dislocations) without any core regularization. Since Holtsmark considered such fields in different versions with different decay laws of the source field, we shall use the generic term "Holtsmark fields" for the random superposition of source fields, each of which decays according to a prescribed power of distance from its source. 

While part of the following is, in essence, a paraphrase of the work of Chandrasekhar and Holtsmark, we nevertheless give all the intermediate steps since this will help to understand our later calculations where we deal with core regularized Holtsmark fields. Also, the excellent though somewhat clumsy work of Holtsmark is not of easy access to readers who do not command the German language, so translating it into modern day English may be of use to the community.  

\subsection{Point like solutes and singular dislocations: the Holtsmark distribution and its variants}

\subsubsection{Solute-dislocation interaction energies and solute-induced displacement fields}

We start with the problem of dislocation-solute interaction energies and evaluate the probability distribution of the random variable
\begin{equation}
	{\cal H} = \sum_{i = 1}^N v_i \frac{z_i}{R_i^3},
\end{equation}
which is proportional to the value of the field ${\cal E}$ in the origin. Alternatively we may consider the same quantity as the value $u_z$ of the $z$ component of the displacement field created by a random arrangement of solutes. The summation runs over all solute coordinate vectors $\Br_i \in V$ and we shall consider the limit $N \to \infty$, $V \to \infty$, $\rho = N/V$ finite. 

To statistically characterize the solute arrangement we introduce the $N$-particle probability density function $w_N(\{\Br_i,v_i\})$ which describes the probability to find the $N$ solutes in the configuration $\{\Br_i,v_i\}$. The solute misfit $v_i$ is taken to be a random variable whose probability density $p(v)$ may allow for discrete values only, e.g. $p(v) = \sum_{\alpha} f_{\alpha} \delta(v-v_{\alpha})$  where the sum is over all defect species and $f_{\alpha}$ are the corresponding atomic fractions. 

The probability density $p({\cal H})$ can then be written as
\begin{equation}
	p({\cal H}) = \int_v \int_{V^N} \delta\left({\cal H} -\sum_{i = 1}^N v_i \frac{z_i}{R_i^3}\right) w_N(\{\Br_i,v_i\}){\rm d}^{3N} r {\rm d}^N v
\end{equation}
where the integral runs over the phase space of all possible $N$-particle configurations and the Dirac $\delta$-function picks only those configurations that yield the field value ${\cal H}$ in the origin. 

We now consider the locations and misfit volumes of the $N$ solutes as independent random variables. Hence, the $N$ particle probability density function $w_N$ factorizes, $w_N = (1/V)^N p(v)^N$. To proceed further, we write the probability distribution $p({\cal H})$ in terms of its Fourier transform $A_{{\cal H}}(Q)$. Exploiting the properties of the Fourier transform of the $\delta$ function and using that the probability density function $w_N$ is $\Br$ independent, we find
\begin{equation}
	p({\cal H}) = \frac{1}{2\pi} \int A_{{\cal H}}(Q) \exp(iQ{\cal H}) {\rm d}Q 
\end{equation}
where 
\begin{equation}
	A_{{\cal H}}(Q) =  \left[\frac{1}{V}\int_V \int_v \exp(-iQ{\cal H}(v,\Br)) {\rm d} ^3 r  p(v) {\rm d} v \right]^N
\end{equation}
In the limit $V,N \to \infty$, $\rho = N/V$ finite we can write
\begin{equation}
	A_{{\cal H}}(Q) =  \lim_{(N,V) \to\infty} \left[1 - \frac{\rho}{N}\int_V p(v) \left(1 - \int_v \exp(-iQ{\cal H}(v,\Br)) {\rm d}^3 r\right)  {\rm d} v \right]^N
\end{equation}
Using the relationship $\exp(x) = \lim_{N \to \infty}(1+x/N)^N$ we can re-write this as
\begin{equation}
	A_{{\cal H}}(Q) = \exp[-\rho C_{{\cal H}}(Q)]\;,\quad C_{{\cal H}}(Q) = \int_V \int_v [1- \exp(Q{\cal H}(v,\Br))]p(v) {\rm d}^3 r  {\rm d} v
 \label{eq:cphi}
\end{equation}
The imaginary part of the integral is zero for symmetry reasons, hence we write the integral in polar coordinates as 
\begin{eqnarray}
C_{{\cal H}}(Q) &=& 2\pi \int_{0}^{\pi}\int_0^{\infty} \int_v \left[1- \cos\left(\frac{Qv \cos \vartheta}{r^2}\right)\right] p(v) r^2 \sin \vartheta \dr {\rm d}\vartheta  {\rm d} v\nonumber\\
&=& 4\pi \int_0^{\pi/2}\int_0^{\infty} \int_v \left[1 - \cos \left(\frac{|Q||v|\cos\vartheta}{r^2}\right)\right] p(v) r^2 \sin \vartheta \dr {\rm d}\vartheta{\rm d} v.
\end{eqnarray}
where in the last step we have used the symmetry of the integrand to replace $v$ and $Q$ with their positively definite moduli $|v|$ and $|Q|$ and to restrict the integration over $\vartheta$ to the domain where $\cos\vartheta$ is positive. This allows us to avoid ambiguity when we now change coordinates by replacing the $r$ integration with an integration over $w = |Q||v|\cos\vartheta/r^2$:
\begin{equation}
	C_{{\cal H}}(Q) = 2\pi |Q|^{3/2} \langle |v|^{3/2} \rangle 
	\int_0^{\infty} [\cos w - 1] w^{-5/2} {\rm d} w \int_0^{\pi/2}(\cos \vartheta)^{3/2} \sin\vartheta {\rm d} \vartheta
\end{equation}
where $\langle |v|^{3/2} \rangle = \int |v|^{3/2} p(v) {\rm d} v$.
Partial integrations give
\begin{equation}
	C_{{\cal H}}(Q) = \frac{16\pi}{15} |Q|^{3/2} \langle |v|^{3/2} \rangle 
	\int_0^{\infty} w^{-1/2} \cos w  {\rm d} w = \frac{4}{15} (2\pi)^{3/2} Q^{3/2} \langle |v|^{3/2} \rangle 
\end{equation}
We now revert to the probability density function $p({\cal H})$ which assumes the form
\begin{equation}
	p({\cal H}) = \frac{1}{2\pi} \int \exp(iQ{\cal H} - (|Q|H_0)^{3/2}) {\rm d}q 
\end{equation}
where the characteristic field $H_0$ is given by
\begin{equation}
	H_0 = 2\pi \left(\frac{4 \langle |v|^{3/2} \rangle \rho }{15}\right)^{2/3}
\end{equation}
Finally we introduce the nondimensional variable $\tilde{{\cal H}} = {\cal H}/H_0$ to write the probability density function in terms of its generating function $F(\tilde{Q})$, $\tilde{Q}= Q H_0$, as 
\begin{equation}
 p(\tilde{{\cal H}}) = \frac{1}{2\pi} \int \exp(i\tilde{Q}\tilde{{\cal H}}) F(\tilde{Q}) {\rm d}\tilde{Q}\;,\quad F(\tilde{Q}) = \exp(-|\tilde{Q}|^{3/2})
\end{equation}
This is the so-called Holtsmark distribution. Its probability density function can be written down in terms of a lengthy combination of hypergeometric functions but there seems little purpose in doing so. However, it is important to study its asymptotics for small and for large values of the random variable ${\cal H}$. At large ${\cal H}$ the distribution, as already shown by Holtsmark, decays like $\tilde{{\cal H}}^{-5/2}$ which puts it into the class of Levy stable distributions and makes it decidedly non Gaussian. The reason for this decay is readily understood when one compares the tail of the distribution with the field generated by the ion closest to the random field point and finds that both distributions asymptotically coincide. In other words, the large-field tail of the Holtsmark distribution is not generated by the superposition of the fields of many independent sources but dominated by the source or sources closest to the point of consideration.  

\subsubsection{Solute-induced shear stresses and Peach-Koehler forces}

We now repeat the calculation for the statistics of solute-induced shear stresses where the relevant random variable has up to prefactors the form
\begin{equation}
	{\cal H} = \sum_{i = 1}^N v_i \frac{x_i z_i}{R_i^5}.
\end{equation}
We follow the same line of argument as above until Eq. (\ref{eq:cphi}) whose counterpart now reads
\begin{equation}
	C_{{\cal H}}(Q) = \int_0^{2\pi} \int_{0}^{\pi}\int_0^{\infty} \int_v \left[1- \cos\left(\frac{Q v \cos \vartheta \sin \vartheta \sin \psi}{r^3}\right)\right] p(v) r^2 \sin \vartheta \dr {\rm d}\vartheta 
	{\rm d}\psi  {\rm d} v.
\end{equation}
Again, we exploit the symmetry of the integrand to replace $v,Q$ with $|v|,|Q|$ and restrict the integration to regions where the angle-dependent function in the argument of the cosine is positive. This gives
\begin{equation}
	C_{{\cal H}}(Q) = 4\int_0^{\pi} \int_{0}^{\pi/2}\int_0^{\infty} \int_v \left[1- \cos\left(\frac{|Q||v| \cos \vartheta \sin \vartheta \sin \psi}{r^3}\right)\right] p(v) r^2 \sin \vartheta \dr {\rm d}\vartheta 
	{\rm d}\psi  {\rm d} v.
\end{equation}
We now substitute $w = (|Q||v|\cos\vartheta\sin\vartheta\sin\psi)/r^3$ to obtain
\begin{eqnarray}
	C_{{\cal H}}(Q) &=& \frac{4|Q|}{3}\int_0^{\infty}  \frac{1- \cos  w}{w^2}{\rm d}w \int_v p(v)|v| {\rm d} v \int_0^{\pi} \int_{0}^{\pi/2}\cos\vartheta\sin^2 \vartheta \sin\psi  {\rm d}\vartheta 
	{\rm d}\psi\nonumber\\
	&=& \frac{8}{9} |Q| \langle |v|\rangle \int_0^{\infty} \frac{\sin w}{w}{\rm d}w = \frac{4\pi}{9} |Q| \langle |v|\rangle .	
\end{eqnarray}
We now again revert to the probability density function $p(\tilde{{\cal H}})$ which assumes the form
\begin{equation}
	p(\tilde{{\cal H}}) = \frac{1}{2\pi} \int \exp(i\tilde{q}\tilde{{\cal H}} - |\tilde{Q}|) {\rm d}\tilde{Q} 
\end{equation}
where $\tilde{Q} = Q  H_0, \tilde{{\cal H}} = {\cal H}/H_0$, and the characteristic field $H_0$ is here given by
\begin{equation}
	H_0 = \frac{4 \pi \langle |v| \rangle \rho }{9}
\end{equation}
In this case, the integration in Eq. (A17) is elementary and the distribution is a Lorentzian:
\begin{equation}
	p(\tilde{{\cal H}}) = \frac{2}{\pi(1 + \tilde{{\cal H}}^2)} 
\end{equation}
Again, the non Gaussian character of this distribution is evident. The tail asymptotics again matches the expectation for a high-field tail that is controlled by the field of the nearest neighbor. 

\subsubsection{Shear stresses created by random dislocation arrangements}

Next, we consider the shear stresses created by an arrangement of randomly located and directed screw dislocations as described in Section \ref{sec:dislocations}. The shear stress field of each dislocation depends only on the coordinates in the perpendicular plane.
The random superposition of the shear stress fields of each 'family' of parallel screw dislocations again corresponds to a Holtsmark field, but this time with $d_{\rm H}=1$ in {\em two} dimensions, as the summation must be carried out over the perpendicular planes.  Due to the stability of the Holtsmark distribution (the sum of Holtsmark distributed variables with the same $\alpha$ is again a similarly distributed variable) and the lack of correlations, the distribution for the entire arrangement is obtained by summation over all 'families'. 

We use without loss of generality the same coordinate system as in the previous calculations, with slip plane normal vector $\Be_z$ and slip vector $\Be_x$. The random shear stress at the origin is according to Eq. \ref{eq:sigxzdis} given by
\begin{equation}
	\sigma_{zx} =  \sum_{i = 1}^N \frac{\mu b}{2\pi} \frac{\eta_{\alpha_i}\cos(\phi_{\alpha}^i)}{R^i_{\alpha}}.
\end{equation}
where the definition of the radius vector $R^i_{\alpha}$ and angle $\phi_{\alpha}^i$ is specific to each 'family' $\alpha$. The corresponding density of screw dislocations in the perpendicular plane is $\rho^{\alpha}$. The factors $\eta_{\alpha}$ are given in Table 5.

The difference with the previous cases resides in the fact that the problem is in effect two-dimensional, and accordingly the densities $\rho_{\alpha} = \rho p_{\alpha}$ of the generating screw dislocations are understood as densities of intersection points per unit area. We obtain
\begin{equation}
	A_{{\cal H}}(Q) = \exp[-\rho C_{{\cal H}}(Q)]\;,\quad C_{{\cal H}}(Q) = \sum_{\alpha} p_{\alpha}\int_V [1- \exp(iQ{\cal H}_{\alpha}(\Br_{\alpha}))]  {\rm d}^2 r. 
 \label{eq:cphistress}
\end{equation}
In cylindrical coordinates:
\begin{equation}
	C_{{\cal H}}(Q) = \sum_{\alpha} p_{\alpha} \int_{-\pi}^{\pi} \int_0^{\infty} \left[1- \cos\left(\frac{\eta_{\alpha}\mu b}{2\pi}\frac{Q \cos \vartheta}{r}\right)\right] r \dr {\rm d}\vartheta
\end{equation}
Substituting $\tilde{r} = 2\pi r/(Q \eta_{\alpha} \mu b)$ we obtain
\begin{eqnarray}
	C_{{\cal H}}(Q) &=&  Q^2 
    \left(\sum_{\alpha} p_{\alpha} \eta_{\alpha}^2\right) \frac{\mu^2 b^2}{4 \pi^2} \int_{0}^{\infty} \int_{0}^{2\pi} \left[1-\cos\left(\left(\frac{1}{\tilde{r}}\cos \vartheta\right)\right)\right]  {\rm d}\vartheta  \tilde{r}{\rm d}\tilde{r} \nonumber\\
    &=& Q^2 \langle 
    \eta_{\alpha}^2\rangle \frac{\mu^2 b^2}{2 \pi} \lim_{X\to\infty} \int_0^X \left[1-J_0\left(\frac{1}{\tilde{r}}\right) \right] \tilde{r} {\rm d}\tilde{r}.
\end{eqnarray}
For finite samples and fully disordered dislocation arrangements, the upper bound of the integral is $X \approx  2\pi R/(Q \eta_{\alpha} \mu b)$ where $R$ indicates the crystal radius.
The integral can be expressed in terms of the generalized hypergeometric function $_2F_3$ (for reference see e.g. \cite{mathai2006generalized}) as follows: 
\begin{equation}
    \int_0^X \left[1-J_0\left(\frac{1}{\tilde{r}}\right) \right] \tilde{r} {\rm d}\tilde{r}  
    = \frac{1}{4} \left(\ln X + 1 - \gamma_{\rm E} + \ln 2\right)
    + \frac{_2F_3(1,1;2,3,3;-1/(4 X^2))}{128 X^2},
\end{equation}
where $\gamma_{\rm E}$ is the Euler-Mascheroni constant. Comparison with numerical evaluations shows that, over a very write range of values of $X$, the term containing the generalized hypergeometric function $_2F_3(\dots)$, which is at leading order proportional to $1/X^2$, can be safely neglected. 

We define a characteristic field $H_0$ by 
\begin{equation}
	H_0 = \mu b \sqrt{\frac{\langle\eta_{\alpha}^2\rangle}{8 \pi}} \sqrt{\rho}.
\end{equation}
Note that this characteristic field has the functional form proposed first by Taylor for the flow stress of a dislocated crystal, which gives a measure of the characteristic magnitude of dislocation interactions. 

We again define non-dimensional fields and conjugate Fourier variables via $\tilde{Q} = Q H_0$ and $\tilde{{\cal H}} = {\cal H}/H_0$. Moreover, we write the upper integration boundary as $X = \chi/\tilde{Q}$ with $\chi = R\sqrt{\rho}$ where $R$ denotes the upper limit of the spatial integration in the original coordinates. In terms of these non-dimensional variables we may write
\begin{equation}
	C_{\phi}(\tilde{Q}) \approx  -  \tilde{Q}^2 \left[ 
    \ln\left( \frac{\chi}{\tilde{Q}}\right) + 1 - \gamma + \ln 2\right].
\end{equation}
where, for a fully random dislocation system, the parameter $\chi$ relates to the system size  via $\chi \approx R \sqrt{32 \pi^3 \rho}$. For correlated dislocation systems we refer to results of \citet{groma1998probability} who demonstrated that, if pair correlations in the dislocation arrangement are taken into account, such correlations may restrict the effective range of interactions to an effective correlation radius $R_{\rm corr}$. We thus set $\chi \approx R_{\rm corr}\sqrt{32\pi^3 \rho}$.
 
With these notations, the probability density function $p(\tilde{\phi})$ assumes the form
\begin{equation}
	p(\tilde{{\cal H}}) = \frac{1}{2\pi} \int \exp\left(i\tilde{Q}\tilde{{\cal H}} + \tilde{Q}^2 \ln \left[\frac{\tilde{Q}}{\chi}\right] \right){\rm d}\tilde{Q} \label{eq:stressdist}
\end{equation}
The behavior of the distribution (\ref{eq:stressdist}) has been studied both in the context of dislocations \citep{groma1998probability,zaiser2002dislocation,beato2005statistical} and in relation to the mathematically equivalent problem of the velocity field of a random arrangement of vortices \citep{chavanis2000statistics}. Moreover, this distribution is experimentally accessible in terms of the X-ray line profiles of dislocated crystals, as the random strain field causes local changes in the spacing of crystal lattice planes and thus a shift of the X-ray lines. In fact, the parameter $\chi$ is essentially identical with the shape parameter $M$ introduced by \citet{wilkens1970determination} for the X-ray line profile. In the asymptotic limit of large fields, the tail of the distribution is given by
\begin{equation}
p(\tilde{{\cal H}}) = \frac{1}{\tilde{{\cal H}}^3}\;,\quad
p(\sigma_{xz}) = \frac{\rho \mu^2 b^2}{8 \pi \sigma_{xz}^3}.     
\end{equation}

\section{Fields of regularized solutes and/or dislocations: core regularized Holtsmark fields}

The large-field asymptotic behavior of the Holtsmark distribution and its variants is directly inherited from the diverging fields as one approaches the field sources -- in our problem, the cores of singular solutes and/or dislocations. We thus now ask how the statistics is changed if we regularize these fields. Given that we consider different regularization methods (regularization of the dislocation stress field via distributed Burgers vector or gradient elasticity) while others have been proposed in the literature (regularization of the solute fields by convolution with a Gaussian, \cite{geslin2021microelasticityI}) we seek a generic formulation which can then be adapted to cover different fields and regularization functions.

\subsubsection{Generic formalism}

We consider a core-regularized Holtsmark field in the form 
\begin{equation}
	{\cal H}_{d,a} = \sum_{i = 1}^N \frac{v_i \eta(\theta,\psi)}{R_i^d} g_d(R_i/a),
\end{equation}
where $v_i$ are source strengths with probability density $p(v)$, $d$ is the decay exponent, $a$ the regularization length, and $\eta$ an angle dependent function which vanishes when integrated over the solid angle $\Omega$. The regularization function $g_d(x)$ has the asymptotic properties
\begin{equation}
\lim_{x\to \infty} g(x) = 1 \quad,\quad \lim_{x\to 0} \frac{g(x)}{x^d} = \frac{1}{a^d} \quad,\int_0^{\infty} [1-g(x)] dx =  \xi.
\end{equation} 
The previously studied examples represent the singular limit cases $\phi_{2,0}$ and $\phi_{3,0}$. 

We proceed as for the singular cases. The generic counterpart of Eqs. (A7)
and (A14) reads
\begin{equation}
	C_{{\cal H}_{a,d}}(Q) = \int_\Omega\int_0^{\infty} \int_v \left[1- \cos\left(\frac{|Q||v||\eta| g_d(r/a)}{r^d}\right)\right] p(v) r^2 \dr {\rm d}\Omega  {\rm d} v.
\end{equation}
At this point, due to the presence of the regularization function $g$ we may simply expand the cosine into a series of powers of its argument, and thus into a series of power of $Q$. We can then carry out the integrations sequentially for each term of the series (note that without $g$, all these integrals would diverge): 
\begin{equation}
	C_{{\cal H}_{a,d}}(Q) = \sum_{k=1}^{\infty} \frac{(-1)^{k+1}}{(2k)!}\frac{ \langle v^{2k} \rangle}{a^{2kd-3}} Q^{2k} \int_\Omega\int_0^{\infty}  \left(\frac{|\eta| g_d(u)}{u^{d}}\right)^{2k} u^2 {\rm d}u {\rm d}\Omega.
\end{equation}
where $u=r/a$. 
We can thus immediately calculate the second moment of the field ${\cal H}$ as
\begin{equation}
	\langle {\cal H}^2 \rangle = -\partial_{Q}^2 A_{{\cal H}}(Q)|_{Q=0} = - \rho 
	\partial_{{\cal H}}^2 C_{{\cal H}}(Q)|_{Q=0} = \frac{\rho \langle v^{2} \rangle}{a^{2d-3}} \int_\Omega\int_0^{\infty}  \left(\frac{|\eta| g_d(u)}{u^{2d-2}}\right)^{2} {\rm d}u {\rm d}\Omega.
\end{equation}
Moreover, it is clear that the tail of the distribution, which is controlled by the small-$Q$ behavior of its characteristic function $A(Q)$, is now asymptotically Gaussian. 

However, the same is not necessarily true for the central part of the probability distribution, which inherits characteristics of the Holtsmark field. To investigate this behavior, we proceed from Eq. (A22) by
expanding the cosine into powers of $1 - g_d$ around $g_d=1$ (corresponding to $r \to \infty$):
\begin{eqnarray}
	C_{{\cal H}_{a,d}}(Q) &=& \int_\Omega\int_0^{\infty} \int_v \left[1- \cos\left(\frac{|Q||v||\eta|}{r^d}\right)\right] p(v) r^2 \dr {\rm d}\Omega  {\rm d} v \nonumber\\
	&=& - \sum_{k=1}^{\infty} \frac{(-1)^k}{(2k-1)!} \int_\Omega\int_0^{\infty} \int_v (q u^d \tilde{g}_d(u))^{2k-1}\sin(q u^d) \frac{a^3}{u^4} p(v)\dr {\rm d}\Omega  {\rm d} v
	\nonumber\\
	&-& \sum_{k=1}^{\infty} \frac{(-1)^k}{(2k)!} \int_\Omega\int_0^{\infty} \int_v (q u^d  \tilde{g}_d(u))^{2k}\cos(qu^d) \frac{a^3}{u^4} p(v)\dr {\rm d}\Omega  {\rm d} v.
	\nonumber\\
\end{eqnarray}
where we have set $u=a/r$ and introduced the notations $q = |Q||v||\eta|/a^d$ and $\tilde{g}_d=1-g_d$. The first term of the series is just the expression for the Holtsmark field. The remaining terms can, in the limit of large $q$, be evaluated using a stationary phase approximation, which demonstrates that for large $q$ these terms are of order $q^{1/d}$. Thus, the large $Q$ asymptotics of $C_{{\cal H}_{a,d}}(Q)$ are dominated by the behavior of the Holtsmark field, $C_{{\cal H}_{a,2}}(Q) \propto |Q|^{3/2}$ and $C_{\phi_{a,3}}(Q) \propto |Q|$. This implies that the {\em central} part of the distribution $p(H_{a,d})$ resembles the distribution of the Holtsmark field, and may not be well approximated by a Gaussian.

\bibliographystyle{cas-model2-names}

\bibliography{references}

\begin{thebibliography}{29}
\expandafter\ifx\csname natexlab\endcsname\relax\def\natexlab#1{#1}\fi
\providecommand{\url}[1]{\texttt{#1}}
\providecommand{\href}[2]{#2}
\providecommand{\path}[1]{#1}
\providecommand{\DOIprefix}{doi:}
\providecommand{\ArXivprefix}{arXiv:}
\providecommand{\URLprefix}{URL: }
\providecommand{\Pubmedprefix}{pmid:}
\providecommand{\doi}[1]{\href{http://dx.doi.org/#1}{\path{#1}}}
\providecommand{\Pubmed}[1]{\href{pmid:#1}{\path{#1}}}
\providecommand{\bibinfo}[2]{#2}
\ifx\xfnm\relax \def\xfnm[#1]{\unskip,\space#1}\fi
\bibitem[{Beato et~al.(2005)Beato, Pietronero and
  Zapperi}]{beato2005statistical}
\bibinfo{author}{Beato, V.}, \bibinfo{author}{Pietronero, L.},
  \bibinfo{author}{Zapperi, S.}, \bibinfo{year}{2005}.
\newblock \bibinfo{title}{Statistical properties of dislocation mutual
  interactions}.
\newblock \bibinfo{journal}{Journal of Statistical Mechanics: Theory and
  Experiment} \bibinfo{volume}{2005}, \bibinfo{pages}{P04011}.
\bibitem[{Cai et~al.(2006)Cai, Arsenlis, Weinberger and Bulatov}]{cai2006non}
\bibinfo{author}{Cai, W.}, \bibinfo{author}{Arsenlis, A.},
  \bibinfo{author}{Weinberger, C.R.}, \bibinfo{author}{Bulatov, V.V.},
  \bibinfo{year}{2006}.
\newblock \bibinfo{title}{A non-singular continuum theory of dislocations}.
\newblock \bibinfo{journal}{Journal of the Mechanics and Physics of Solids}
  \bibinfo{volume}{54}, \bibinfo{pages}{561--587}.
\bibitem[{Chandrasekhar(1943)}]{chandrasekhar1943stochastic}
\bibinfo{author}{Chandrasekhar, S.}, \bibinfo{year}{1943}.
\newblock \bibinfo{title}{Stochastic problems in physics and astronomy}.
\newblock \bibinfo{journal}{Reviews of modern physics} \bibinfo{volume}{15},
  \bibinfo{pages}{1}.
\bibitem[{Chavanis and Sire(2000)}]{chavanis2000statistics}
\bibinfo{author}{Chavanis, P.H.}, \bibinfo{author}{Sire, C.},
  \bibinfo{year}{2000}.
\newblock \bibinfo{title}{Statistics of velocity fluctuations arising from a
  random distribution of point vortices: The speed of fluctuations and the
  diffusion coefficient}.
\newblock \bibinfo{journal}{Physical Review E} \bibinfo{volume}{62},
  \bibinfo{pages}{490}.
\bibitem[{Eshelby(1956)}]{eshelby1956continuum}
\bibinfo{author}{Eshelby, J.}, \bibinfo{year}{1956}.
\newblock \bibinfo{title}{The continuum theory of lattice defects}.
\newblock \bibinfo{journal}{Solid state physics} \bibinfo{volume}{3},
  \bibinfo{pages}{79--144}.
\bibitem[{Fan et~al.(2021)Fan, Wang, El-Awady, Raabe and
  Zaiser}]{fan2021strain}
\bibinfo{author}{Fan, H.}, \bibinfo{author}{Wang, Q.},
  \bibinfo{author}{El-Awady, J.A.}, \bibinfo{author}{Raabe, D.},
  \bibinfo{author}{Zaiser, M.}, \bibinfo{year}{2021}.
\newblock \bibinfo{title}{Strain rate dependency of dislocation plasticity}.
\newblock \bibinfo{journal}{Nature communications} \bibinfo{volume}{12},
  \bibinfo{pages}{1845}.
\bibitem[{Geslin et~al.(2021)Geslin, Rida and
  Rodney}]{geslin2021microelasticityII}
\bibinfo{author}{Geslin, P.A.}, \bibinfo{author}{Rida, A.},
  \bibinfo{author}{Rodney, D.}, \bibinfo{year}{2021}.
\newblock \bibinfo{title}{Microelasticity model of random alloys. part ii:
  displacement and stress correlations}.
\newblock \bibinfo{journal}{Journal of the Mechanics and Physics of Solids}
  \bibinfo{volume}{153}, \bibinfo{pages}{104480}.
\bibitem[{Geslin and Rodney(2021)}]{geslin2021microelasticityI}
\bibinfo{author}{Geslin, P.A.}, \bibinfo{author}{Rodney, D.},
  \bibinfo{year}{2021}.
\newblock \bibinfo{title}{Microelasticity model of random alloys. part i: mean
  square displacements and stresses}.
\newblock \bibinfo{journal}{Journal of the Mechanics and Physics of Solids}
  \bibinfo{volume}{153}, \bibinfo{pages}{104479}.
\bibitem[{Groma and Bak{\'o}(1998)}]{groma1998probability}
\bibinfo{author}{Groma, I.}, \bibinfo{author}{Bak{\'o}, B.},
  \bibinfo{year}{1998}.
\newblock \bibinfo{title}{Probability distribution of internal stresses in
  parallel straight dislocation systems}.
\newblock \bibinfo{journal}{Physical Review B} \bibinfo{volume}{58},
  \bibinfo{pages}{2969}.
\bibitem[{Groma and Bak{\'o}(2000)}]{groma2000dislocation}
\bibinfo{author}{Groma, I.}, \bibinfo{author}{Bak{\'o}, B.},
  \bibinfo{year}{2000}.
\newblock \bibinfo{title}{Dislocation patterning: from micro-to mesoscale
  description}.
\newblock \bibinfo{journal}{Physical review letters} \bibinfo{volume}{84},
  \bibinfo{pages}{1487}.
\bibitem[{Holtsmark(1919)}]{holtsmark1919verbreiterung}
\bibinfo{author}{Holtsmark, J.}, \bibinfo{year}{1919}.
\newblock \bibinfo{title}{{\"U}ber die verbreiterung von spektrallinien}.
\newblock \bibinfo{journal}{Annalen der Physik} \bibinfo{volume}{363},
  \bibinfo{pages}{577--630}.
\bibitem[{Kuvshinov and Schep(2000)}]{kuvshinov2000holtsmark}
\bibinfo{author}{Kuvshinov, B.N.}, \bibinfo{author}{Schep, T.J.},
  \bibinfo{year}{2000}.
\newblock \bibinfo{title}{Holtsmark distributions in point-vortex systems}.
\newblock \bibinfo{journal}{Physical review letters} \bibinfo{volume}{84},
  \bibinfo{pages}{650}.
\bibitem[{LaRosa et~al.(2019)LaRosa, Shih, Varvenne and
  Ghazisaeidi}]{larosa2019solid}
\bibinfo{author}{LaRosa, C.R.}, \bibinfo{author}{Shih, M.},
  \bibinfo{author}{Varvenne, C.}, \bibinfo{author}{Ghazisaeidi, M.},
  \bibinfo{year}{2019}.
\newblock \bibinfo{title}{Solid solution strengthening theories of high-entropy
  alloys}.
\newblock \bibinfo{journal}{Materials Characterization} \bibinfo{volume}{151},
  \bibinfo{pages}{310--317}.
\bibitem[{Lazar(2013)}]{lazar2013fundamentals}
\bibinfo{author}{Lazar, M.}, \bibinfo{year}{2013}.
\newblock \bibinfo{title}{The fundamentals of non-singular dislocations in the
  theory of gradient elasticity: Dislocation loops and straight dislocations}.
\newblock \bibinfo{journal}{International Journal of Solids and Structures}
  \bibinfo{volume}{50}, \bibinfo{pages}{352--362}.
\bibitem[{Lazar(2019)}]{lazar2019non}
\bibinfo{author}{Lazar, M.}, \bibinfo{year}{2019}.
\newblock \bibinfo{title}{A non-singular continuum theory of point defects
  using gradient elasticity of bi-helmholtz type}.
\newblock \bibinfo{journal}{Philosophical Magazine} \bibinfo{volume}{99},
  \bibinfo{pages}{1563--1601}.
\bibitem[{Mathai and Saxena(2006)}]{mathai2006generalized}
\bibinfo{author}{Mathai, A.M.}, \bibinfo{author}{Saxena, R.K.},
  \bibinfo{year}{2006}.
\newblock \bibinfo{title}{Generalized hypergeometric functions with
  applications in statistics and physical sciences}. volume
  \bibinfo{volume}{348}.
\newblock \bibinfo{publisher}{Springer}.
\bibitem[{Nattermann et~al.(1992)Nattermann, Stepanow, Tang and
  Leschhorn}]{nattermann1992dynamics}
\bibinfo{author}{Nattermann, T.}, \bibinfo{author}{Stepanow, S.},
  \bibinfo{author}{Tang, L.H.}, \bibinfo{author}{Leschhorn, H.},
  \bibinfo{year}{1992}.
\newblock \bibinfo{title}{Dynamics of interface depinning in a disordered
  medium}.
\newblock \bibinfo{journal}{Journal de Physique II} \bibinfo{volume}{2},
  \bibinfo{pages}{1483--1488}.
\bibitem[{Po et~al.(2014)Po, Lazar, Seif and Ghoniem}]{po2014singularity}
\bibinfo{author}{Po, G.}, \bibinfo{author}{Lazar, M.}, \bibinfo{author}{Seif,
  D.}, \bibinfo{author}{Ghoniem, N.}, \bibinfo{year}{2014}.
\newblock \bibinfo{title}{Singularity-free dislocation dynamics with strain
  gradient elasticity}.
\newblock \bibinfo{journal}{Journal of the Mechanics and Physics of Solids}
  \bibinfo{volume}{68}, \bibinfo{pages}{161--178}.
\bibitem[{Rida et~al.(2022)Rida, Martinez, Rodney and
  Geslin}]{rida2022influence}
\bibinfo{author}{Rida, A.}, \bibinfo{author}{Martinez, E.},
  \bibinfo{author}{Rodney, D.}, \bibinfo{author}{Geslin, P.A.},
  \bibinfo{year}{2022}.
\newblock \bibinfo{title}{Influence of stress correlations on dislocation glide
  in random alloys}.
\newblock \bibinfo{journal}{Physical Review Materials} \bibinfo{volume}{6},
  \bibinfo{pages}{033605}.
\bibitem[{Rodney et~al.(2024)Rodney, Geslin, Patinet, D{\'e}mery and
  Rosso}]{rodney2024does}
\bibinfo{author}{Rodney, D.}, \bibinfo{author}{Geslin, P.A.},
  \bibinfo{author}{Patinet, S.}, \bibinfo{author}{D{\'e}mery, V.},
  \bibinfo{author}{Rosso, A.}, \bibinfo{year}{2024}.
\newblock \bibinfo{title}{Does the larkin length exist?}
\newblock \bibinfo{journal}{Modelling and Simulation in Materials Science and
  Engineering} \bibinfo{volume}{32}, \bibinfo{pages}{035007}.
\bibitem[{Saada and Sornette(1995)}]{saada1995long}
\bibinfo{author}{Saada, G.}, \bibinfo{author}{Sornette, D.},
  \bibinfo{year}{1995}.
\newblock \bibinfo{title}{Long-range stress field fluctuations induced by
  random dislocation arrays: a unified spectral approach}.
\newblock \bibinfo{journal}{Acta metallurgica et materialia}
  \bibinfo{volume}{43}, \bibinfo{pages}{313--318}.
\bibitem[{Varvenne et~al.(2017)Varvenne, Leyson, Ghazisaeidi and
  Curtin}]{Varvenne2017solute}
\bibinfo{author}{Varvenne, C.}, \bibinfo{author}{Leyson, G.},
  \bibinfo{author}{Ghazisaeidi, M.}, \bibinfo{author}{Curtin, W.},
  \bibinfo{year}{2017}.
\newblock \bibinfo{title}{Solute strengthening in random alloys}.
\newblock \bibinfo{journal}{Acta Materialia} \bibinfo{volume}{124},
  \bibinfo{pages}{660--683}.
\bibitem[{Varvenne et~al.(2016)Varvenne, Luque and Curtin}]{Varvenne2016theory}
\bibinfo{author}{Varvenne, C.}, \bibinfo{author}{Luque, A.},
  \bibinfo{author}{Curtin, W.A.}, \bibinfo{year}{2016}.
\newblock \bibinfo{title}{Theory of strengthening in fcc high entropy alloys}.
\newblock \bibinfo{journal}{Acta Materialia} \bibinfo{volume}{118},
  \bibinfo{pages}{164--176}.
\bibitem[{Wilkens(1970)}]{wilkens1970determination}
\bibinfo{author}{Wilkens, M.}, \bibinfo{year}{1970}.
\newblock \bibinfo{title}{The determination of density and distribution of
  dislocations in deformed single crystals from broadened x-ray diffraction
  profiles}.
\newblock \bibinfo{journal}{Physica status solidi (a)} \bibinfo{volume}{2},
  \bibinfo{pages}{359--370}.
\bibitem[{Zaiser(2002)}]{zaiser2002dislocation}
\bibinfo{author}{Zaiser, M.}, \bibinfo{year}{2002}.
\newblock \bibinfo{title}{Dislocation motion in a random solid solution}.
\newblock \bibinfo{journal}{Philosophical Magazine A} \bibinfo{volume}{82},
  \bibinfo{pages}{2869--2883}.
\bibitem[{Zaiser and Sandfeld(2014)}]{zaiser2014scaling}
\bibinfo{author}{Zaiser, M.}, \bibinfo{author}{Sandfeld, S.},
  \bibinfo{year}{2014}.
\newblock \bibinfo{title}{Scaling properties of dislocation simulations in the
  similitude regime}.
\newblock \bibinfo{journal}{Modelling and Simulation in Materials Science and
  Engineering} \bibinfo{volume}{22}, \bibinfo{pages}{065012}.
\bibitem[{Zaiser and Wu(2022)}]{zaiser2022pinning}
\bibinfo{author}{Zaiser, M.}, \bibinfo{author}{Wu, R.}, \bibinfo{year}{2022}.
\newblock \bibinfo{title}{Pinning of dislocations in disordered alloys: effects
  of dislocation orientation}.
\newblock \bibinfo{journal}{Materials Theory} \bibinfo{volume}{6},
  \bibinfo{pages}{1--13}.
\bibitem[{Zbib et~al.(1998)Zbib, Rhee and Hirth}]{Zbib1998_IJMS}
\bibinfo{author}{Zbib, H.M.}, \bibinfo{author}{Rhee, M.},
  \bibinfo{author}{Hirth, J.P.}, \bibinfo{year}{1998}.
\newblock \bibinfo{title}{On plastic deformation and the dynamics of 3d
  dislocations}.
\newblock \bibinfo{journal}{International Journal of Mechanical Sciences}
  \bibinfo{volume}{40}, \bibinfo{pages}{113--127}.
\bibitem[{Zbib and de~la Rubia(2002)}]{zbib2002multiscale}
\bibinfo{author}{Zbib, H.M.}, \bibinfo{author}{de~la Rubia, T.D.},
  \bibinfo{year}{2002}.
\newblock \bibinfo{title}{A multiscale model of plasticity}.
\newblock \bibinfo{journal}{International Journal of Plasticity}
  \bibinfo{volume}{18}, \bibinfo{pages}{1133--1163}.

\end{thebibliography}


\end{document}